\documentclass[journal=jctcce,manuscript=article]{achemso}
\usepackage{amssymb, amsmath, amsfonts, multicol, multirow, longtable, array, mathpazo}
\usepackage{amsmath}
\usepackage{lscape}
\usepackage[version=3]{mhchem} % Formula subscripts using \ce{}
\usepackage{appendix}
\usepackage{soul} %text highlighting
\usepackage[usenames]{color}
\usepackage{makecell}
\usepackage{enumerate}
\usepackage{appendix}
\usepackage{xr}
\usepackage[T1]{fontenc} % Use modern font encodingss
\usepackage{booktabs}
\usepackage{tikz}
\usepackage{threeparttable}
\usepackage{threeparttablex}
\usepackage{graphicx}
\usepackage{subfigure}
\usepackage{colortbl}
\usepackage{framed}
\usepackage{verbatim}
\usepackage{amsbsy}
\usepackage{bm}% bold math
\usepackage{bbm}
\usepackage[normalem]{ulem}
\usepackage{float}
\usepackage[linesnumbered,boxed]{algorithm2e}
\usepackage{algorithm2e}
\usepackage{amsmath}
\usepackage{diagbox}
\usepackage{mciteplus}
\newcounter{xscheme}

\newfloat{Algorithm}{htbp}{alg}
\floatname{Algorithm}{Algorithm}

\newcounter{exe}[figure]
\newcommand{\iexe}{\refstepcounter{exe}\the\value{exe}:}

\newcommand{\CSMGGSnum}{
	\citenum{DDB_131,DDB_132,DDB_134,MDB06,MDB2012,MDB2015,MDB2019,DDB22,GMTKN24,GMTKN30,GMTKN55,MGCDB84,DDB_90,DDB_45,G2_ion,W4-11,DDB_130,DDB_152,G1,G2-91,DDB_88,FCIQMC-diatom,FCI21,DDB_176,DDB_35,DDB_39,DDB_40,DDB_41,DDB_47,DDB_54,DDB_61,DDB_75,DDB_78,BSR36-2009,DDB_108,DDB_138,DDB_139,DDB_148,DDB_151,DDB_157,DDB_163,DDB_164,DDB_165,DDB_169,DDB_170,DDB_173,DDB_225,DDB_43,DDB_52,DDB_57,DDB_59,DDB_65,DDB_67,DDB_71,DDB_76,DDB_77,DDB_79,DDB_80,DDB_81,DDB_82,DDB_85,DDB_105,DDB_109,DDB_135,DDB_136,DDB_137,DDB_141,DDB_142,DDB_144,DDB_155,DDB_156,DDB_159,DDB_160,DDB_161,DDB_162,DDB_166,DDB_167,DDB_168,DDB_171,DDB_172,DDB_174,DDB_226,DDB_48,DDB_49,DDB_53,DDB_56,DDB_58,DDB_63,DDB_64,DDB_69,DDB_72,DDB_74,DDB_83,DDB_104,DDB_106,S22,DDB_140,DDB_145,S66,DDB_110,DDB_154}
}

\newcommand{\OSMGGSnum}{
	\citenum{DDB_131,DDB_132,DDB_134,MDB06,MDB2012,MDB2015,MDB2019,DDB22,GMTKN24,GMTKN30,GMTKN55,MGCDB84,DDB_90,DDB_45,G2_ion,W4-11,DDB_130,DDB_152,G1,G2-91,DDB_88,FCIQMC-diatom,FCI21,DDB_176,DDB_66}
}

\newcommand{\CSTMGSnum}{
	\citenum{DDB_131,DDB_132,DDB_134,MDB06,MDB2012,MDB2015,MDB2019,DDB22,ccCA-TM_11,3dBE70,DDB_127,TMC151,CUAGAU,CUAGAU-2,Ln54,An66,ccCA-TM_11all,DDB_92,3dMLBE20,TMHoFangela,TMdissocEneMR2016,AnIPAngela2022,LnSLnSe,LnIP4,DDB_112,DDB_113,DDB_114,DDB_115}
}

\newcommand{\OSTMGSnum}{
	\citenum{DDB_131,DDB_132,DDB_134,MDB06,MDB2012,MDB2015,MDB2019,DDB22,ccCA-TM_11,3dBE70,DDB_127,TMC151,CUAGAU,CUAGAU-2,Ln54,An66,ccCA-TM_11all,DDB_92,3dMLBE20,TMHoFangela,TMdissocEneMR2016,AnIPAngela2022,LnSLnSe,LnIP4,DDB_227}
}

\newcommand{\CSMGESnum}{
	\citenum{MDB06,MDB2012,MDB2015,MDB2019,DDB22,send_adiabatic,DDB_100,QUEST4,DDB_116,benchmarkSDS,DDB_118,DDB_123,34small2007Jacquemin,jacquemin2008td,483org2009Jacquemin,40dyes2012Jacquemin,parac2002comparison,5OrgMol2009Grimme,40large2004Grimme,2004Boeij,76_S-dye2001fabian,semiemp2008matsuura,6org3small2008Herbert,XABOOM2021,15OrgMol2013Mennucci,shao2019benchmarking,grabarz2021benchmarking,bogo2024benchmarking,loos0-0_18,13BODIPY2022Helal,DDB_119,DDB_120,RLex80_2015Jacquemin,benchmarkCC_DFT13,12OrgMol2010Grimme,Conical2013Filatov,gronowski2017td,mewes2018benchmarking,V_0_Adiabatic,PAH2009Sauer,kozma2020new,loos0-0_19JCTC,QUEST1,QUEST3,QUEST5,QUEST6,QUEST7,DDB_122,Tozer2008,polyenes2008Marian,17BODIPY2015Brown,yang2016nature,budzak2017accurate,spezia2017excited,wiebeler2017excitation,shi2019excited,biradical2017,QUEST2,Thiel08,scott2019multireference}
}

\newcommand{\OSMGESnum}{
	\citenum{MDB06,MDB2012,MDB2015,MDB2019,DDB22,send_adiabatic,DDB_100,QUEST4,QUESTsatellite,DDB_116,benchmarkSDS,07GrimmeNeese,LZD1,LZD2,openshellGordon}
}

\newcommand{\CSTMESnum}{
	\citenum{MDB06,MDB2012,MDB2015,MDB2019,DDB22,send_adiabatic,radon2021spin,aoto2017arrive,7TMcomplex2014Thiel,huntington2015application,QUEST8}
}

\newcommand{\OSTMESnum}{
	\citenum{MDB06,MDB2012,MDB2015,MDB2019,DDB22,send_adiabatic,radon2021spin,aoto2017arrive,SUO3,radon2019benchmarking,khedkar2021modern}
}

\newcommand{\CSMGGS}{
	\cite{DDB_131,DDB_132,DDB_134,MDB06,MDB2012,MDB2015,MDB2019,DDB22,GMTKN24,GMTKN30,GMTKN55,MGCDB84,DDB_90,DDB_45,G2_ion,W4-11,DDB_130,DDB_152,G1,G2-91,DDB_88,FCIQMC-diatom,FCI21,DDB_176,DDB_35,DDB_39,DDB_40,DDB_41,DDB_47,DDB_54,DDB_61,DDB_75,DDB_78,BSR36-2009,DDB_108,DDB_138,DDB_139,DDB_148,DDB_151,DDB_157,DDB_163,DDB_164,DDB_165,DDB_169,DDB_170,DDB_173,DDB_225,DDB_43,DDB_52,DDB_57,DDB_59,DDB_65,DDB_67,DDB_71,DDB_76,DDB_77,DDB_79,DDB_80,DDB_81,DDB_82,DDB_85,DDB_105,DDB_109,DDB_135,DDB_136,DDB_137,DDB_141,DDB_142,DDB_144,DDB_155,DDB_156,DDB_159,DDB_160,DDB_161,DDB_162,DDB_166,DDB_167,DDB_168,DDB_171,DDB_172,DDB_174,DDB_226,DDB_48,DDB_49,DDB_53,DDB_56,DDB_58,DDB_63,DDB_64,DDB_69,DDB_72,DDB_74,DDB_83,DDB_104,DDB_106,S22,DDB_140,DDB_145,S66,DDB_110,DDB_154}
}

\newcommand{\OSMGGS}{
	\cite{DDB_131,DDB_132,DDB_134,MDB06,MDB2012,MDB2015,MDB2019,DDB22,GMTKN24,GMTKN30,GMTKN55,MGCDB84,DDB_90,DDB_45,G2_ion,W4-11,DDB_130,DDB_152,G1,G2-91,DDB_88,FCIQMC-diatom,FCI21,DDB_176,DDB_66}
}

\newcommand{\CSOSTMGS}{
	\cite{DDB_131,DDB_132,DDB_134,MDB06,MDB2012,MDB2015,MDB2019,DDB22,ccCA-TM_11,3dBE70,DDB_127,TMC151,CUAGAU,CUAGAU-2,Ln54,An66,ccCA-TM_11all,DDB_92,3dMLBE20,TMHoFangela,TMdissocEneMR2016,AnIPAngela2022,LnSLnSe,LnIP4,DDB_112,DDB_113,DDB_114,DDB_115,DDB_227}
}

\newcommand{\GSSRuse}{
	\cite{DDB_131,DDB_132,DDB_134,MDB06,MDB2012,MDB2015,MDB2019,DDB22,GMTKN24,GMTKN30,GMTKN55,MGCDB84,DDB_90,DDB_45,G2_ion,W4-11,DDB_130,DDB_152,G1,G2-91,DDB_88,ccCA-TM_11,3dBE70,DDB_127,TMC151,CUAGAU,CUAGAU-2,Ln54,An66,ccCA-TM_11all, DDB_92, 3dMLBE20,TMHoFangela,TMdissocEneMR2016,AnIPAngela2022,LnSLnSe,LnIP4,DDB_35,DDB_39,DDB_40,DDB_41,DDB_47,DDB_54,DDB_61,DDB_75,DDB_78,BSR36-2009,DDB_108,DDB_138,DDB_139,DDB_148,DDB_151,DDB_157,DDB_163,DDB_164,DDB_165,DDB_169,DDB_170,DDB_173,DDB_225,DDB_43,DDB_52,DDB_57,DDB_59,DDB_65,DDB_67,DDB_71,DDB_76,DDB_77,DDB_79,DDB_80,DDB_81,DDB_82,DDB_85,DDB_105,DDB_109,DDB_135,DDB_136,DDB_137,DDB_141,DDB_142,DDB_144,DDB_155,DDB_156,DDB_159,DDB_160,DDB_161,DDB_162,DDB_166,DDB_167,DDB_168,DDB_171,DDB_172,DDB_174,DDB_226,DDB_48,DDB_49,DDB_53,DDB_56,DDB_58,DDB_63,DDB_64,DDB_69,DDB_72,DDB_74,DDB_83,DDB_104,DDB_106,S22,DDB_140,DDB_145,S66,DDB_110,DDB_154,DDB_66,DDB_227,DDB_50,DDB_60,DDB_147,DDB_150,DDB_42,DDB_44,DDB_55,DDB_62,DDB_70,DDB_73,G2-97,BSR36-2010,DDB_101,DDB_102,DDB_146,DDB_175}
}

\newcommand{\GSDFTuse}{
	\cite{DDB_131,DDB_132,DDB_134,MDB06,MDB2012,MDB2015,MDB2019,DDB22,GMTKN24,GMTKN30,GMTKN55,MGCDB84,DDB_90,DDB_45,G2_ion,W4-11,DDB_130,DDB_152,ccCA-TM_11,3dBE70,DDB_127,TMC151,CUAGAU,CUAGAU-2,Ln54,An66,ccCA-TM_11all,DDB_92,3dMLBE20,DDB_35,DDB_39,DDB_40,DDB_41,DDB_47,DDB_54,DDB_61,DDB_75,DDB_78,BSR36-2009,DDB_108,DDB_138,DDB_139,DDB_148,DDB_151,DDB_157,DDB_163,DDB_164,DDB_165,DDB_169,DDB_170,DDB_173,DDB_225,DDB_43,DDB_52,DDB_57,DDB_59,DDB_65,DDB_67,DDB_71,DDB_76,DDB_77,DDB_79,DDB_80,DDB_81,DDB_82,DDB_85,DDB_105,DDB_109,DDB_135,DDB_136,DDB_137,DDB_141,DDB_142,DDB_144,DDB_155,DDB_156,DDB_159,DDB_160,DDB_161,DDB_162,DDB_166,DDB_167,DDB_168,DDB_171,DDB_172,DDB_174,DDB_226}
}

\newcommand{\CSMGES}{
	\cite{MDB06,MDB2012,MDB2015,MDB2019,DDB22,send_adiabatic,DDB_100,QUEST4,DDB_116,benchmarkSDS,DDB_118,DDB_123,34small2007Jacquemin,jacquemin2008td,483org2009Jacquemin,40dyes2012Jacquemin,parac2002comparison,5OrgMol2009Grimme,40large2004Grimme,2004Boeij,76_S-dye2001fabian,semiemp2008matsuura,6org3small2008Herbert,XABOOM2021,15OrgMol2013Mennucci,shao2019benchmarking,grabarz2021benchmarking,bogo2024benchmarking,loos0-0_18,13BODIPY2022Helal,DDB_119,DDB_120,RLex80_2015Jacquemin,benchmarkCC_DFT13,12OrgMol2010Grimme,Conical2013Filatov,gronowski2017td,mewes2018benchmarking,V_0_Adiabatic,PAH2009Sauer,kozma2020new,loos0-0_19JCTC,QUEST1,QUEST3,QUEST5,QUEST6,QUEST7,DDB_122,Tozer2008,polyenes2008Marian,17BODIPY2015Brown,yang2016nature,budzak2017accurate,spezia2017excited,wiebeler2017excitation,shi2019excited,biradical2017,QUEST2,Thiel08,scott2019multireference}
}

\newcommand{\OSMGES}{
	\cite{MDB06,MDB2012,MDB2015,MDB2019,DDB22,send_adiabatic,DDB_100,QUEST4,QUESTsatellite,DDB_116,benchmarkSDS,07GrimmeNeese,LZD1,LZD2,openshellGordon}
}

\newcommand{\CSTMES}{
	\cite{MDB06,MDB2012,MDB2015,MDB2019,DDB22,send_adiabatic,radon2021spin,aoto2017arrive,7TMcomplex2014Thiel,huntington2015application,QUEST8}
}

\newcommand{\OSTMES}{
	\cite{MDB06,MDB2012,MDB2015,MDB2019,DDB22,send_adiabatic,radon2021spin,aoto2017arrive,SUO3,radon2019benchmarking,khedkar2021modern}
}

\newcommand{\ESSR}{
	\cite{MDB06,MDB2012,MDB2015,MDB2019,DDB22,send_adiabatic,DDB_100,QUEST4,DDB_116,radon2021spin,aoto2017arrive,DDB_118,DDB_123,34small2007Jacquemin,jacquemin2008td,483org2009Jacquemin,40dyes2012Jacquemin,parac2002comparison,5OrgMol2009Grimme,40large2004Grimme,2004Boeij,76_S-dye2001fabian,semiemp2008matsuura,6org3small2008Herbert,XABOOM2021,15OrgMol2013Mennucci,shao2019benchmarking,grabarz2021benchmarking,bogo2024benchmarking,loos0-0_18,13BODIPY2022Helal,DDB_119,DDB_120,RLex80_2015Jacquemin,benchmarkCC_DFT13,12OrgMol2010Grimme,Conical2013Filatov,gronowski2017td,mewes2018benchmarking,V_0_Adiabatic,PAH2009Sauer,kozma2020new,loos0-0_19JCTC,QUEST1,QUEST3,QUEST5,QUEST6,QUEST7,QUESTsatellite,DDB_122,Tozer2008,polyenes2008Marian,17BODIPY2015Brown,yang2016nature,budzak2017accurate,spezia2017excited,wiebeler2017excitation,shi2019excited,biradical2017,QUEST2,Thiel08,07GrimmeNeese,LZD1,LZD2,openshellGordon,7TMcomplex2014Thiel,huntington2015application,QUEST8,SUO3,radon2019benchmarking,DDB_121,benchmark_semi,QUEST4_TDDHDF_2022,QUEST_CSF_CI,ADC14,ThielCC13,ThielCC14,ThielCC17, silva2008benchmarks,rohrdanz2009long, jacquemin2010assessment, jacquemin2010performances, mardirossian2011benchmark, jacquemin2011assessment, huix2011assessment, della2011accurate, trani2011time, peverati2012performance, maier2016validation,sauer2015performance,ADC14,yang2014excitation,sauer2009benchmarks, demel2013additional, piecuch2015benchmarking, tajti2016investigation, rishi2017excited, dutta2018exploring,QUEST2_extension,QUEST_CC4}
}

\newcommand{\ESMR}{
	\cite{DDB_116,benchmarkSDS,radon2021spin,aoto2017arrive,DDB_122,Tozer2008,polyenes2008Marian,17BODIPY2015Brown,yang2016nature,budzak2017accurate,spezia2017excited,wiebeler2017excitation,shi2019excited,biradical2017,QUEST2,Thiel08,scott2019multireference,7TMcomplex2014Thiel,huntington2015application,QUEST8,radon2019benchmarking,khedkar2021modern,QUEST2_extension,DDB_125,QUEST1and3,NEVPT2benchmark,CASPT2basis,Thiel-CASSCF,Thiel-MREOMCC,Thiel-icMRCC2,Thiel-RegularizedCASPT2,Thiel-extendedCASPT2,Thiel-MKMRCC-geom,Thiel-AC,QUEST_CASPT3,QUEST_SHCI_rad}
}

\setkeys{acs}{maxauthors = 0} % references of many authors

\author{Yangyang Song}
\affiliation{Qingdao Institute for Theoretical and Computational Sciences, School of Chemistry and Chemical Engineering, Shandong University, Qingdao 266237, China}
\author{Ning Zhang}
\affiliation{Qingdao Institute for Theoretical and Computational Sciences, School of Chemistry and Chemical Engineering, Shandong University, Qingdao 266237, China}
\author{Yibo Lei}
\affiliation{Key Laboratory of Synthetic and Natural Functional Molecule of the Ministry of Education, College of Chemistry \& Materials Science, Shaanxi key Laboratory of Physico-Inorganic Chemistry, Northwest University, Xi'an 710127, China}
\author{Yang Guo}
\affiliation{Qingdao Institute for Theoretical and Computational Sciences, School of Chemistry and Chemical Engineering, Shandong University, Qingdao 266237, China}
\author{Wenjian Liu}\email{liuwj@sdu.edu.cn}
\affiliation{Qingdao Institute for Theoretical and Computational Sciences, School of Chemistry and Chemical Engineering, Shandong University, Qingdao 266237, China}

\title{QUEST\#4X: an extension of QUEST\#4 for benchmarking multireference wavefunction methods}

\setkeys{acs}{articletitle=true}

\begin{document}

\newpage

\begin{abstract}
Given a number of datasets for evaluating the performance of
single reference methods for the low-lying excited states of closed-shell molecules,
a comprehensive dataset for assessing the performance of multireference methods for the low-lying excited states of open-shell systems is
still desired. For this reason, we propose an extension (QUEST\#4X) of the radical subset of QUEST\#4 [J. Chem. Theory Comput. 2020, 16, 3720] to cover
110 doublet and 39 quartet excited states.
Near-exact results obtained by iCIPT2 (iterative configuration interaction with selection and second-order
perturbation correction) are taken as benchmark to calibrate SDSCI (static-dynamic-static configuration interaction)
and SDSPT2 (static-dynamic-static second-order perturbation theory),
which are minimal MRCI and CI-like perturbation theory, respectively. It is found that SDSCI is very close in accuracy
to ic-MRCISD (internally contracted multireference configuration interaction with singles and doubles),
although its computational cost is just that of one iteration of the latter.
Unlike most variants of MRPT2, SDSPT2 treats single and multiple states in the same way, and performs similarly as MS-NEVPT2 (multi-state
n-electron valence second-order perturbation theory). These findings put SDSCI and SDSPT2 on
a firm basis.

\end{abstract}

\maketitle
\clearpage
\newpage

\section{Introduction}
\begin{figure}
	\centering
{\resizebox{0.7\textwidth}{!}{\includegraphics{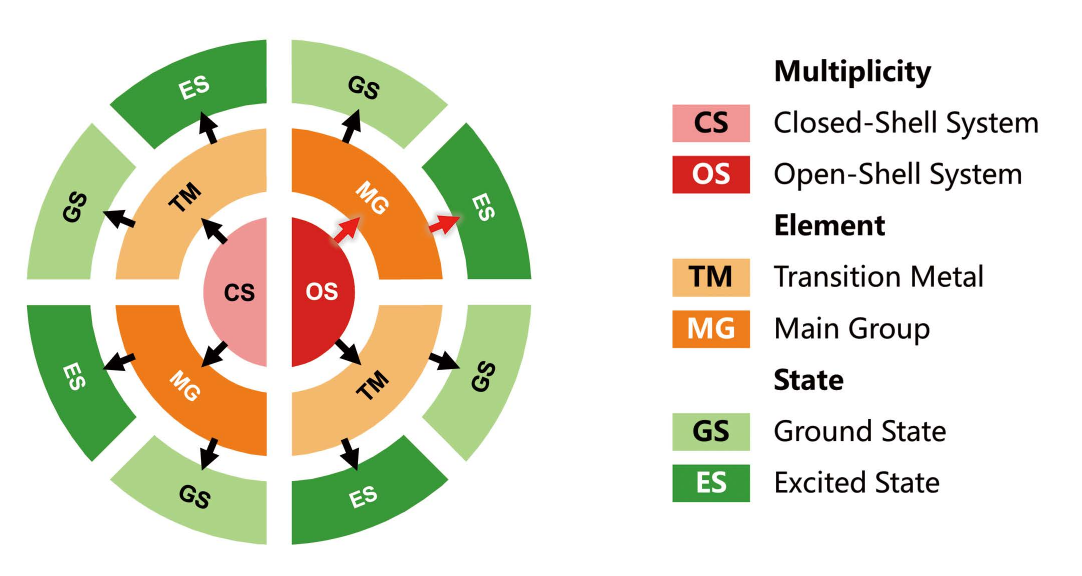}}}
         \caption{Classification of datasets}
	\label{fig-benchmark-type}
\end{figure}

The last decades have witnessed fast progresses in the development of both wavefunction- and density-based quantum chemical methods
for describing electronic structures of chemical systems. It is generally true that the strengths and weaknesses
of each of such methods should be uncovered before they can be applied safely to unknown problems. To this end,
some standardized datasets should be established, such that they can be employed to identify the error bars of a given method.
A dataset is composed of two ingredients, target systems and reference data. The former refers to chosen molecules and their properties,
whereas the latter refers to corresponding experimental or highly accurate theoretical values. The target systems can be
classified according to the simple criteria,
(1) closed-shell (CS) or open-shell (OS), (2) main group (MG) or transition metal (TM),
(3) ground state (GS) or excited state (ES), thereby leading to 8 types of datasets
(see Fig. \ref{fig-benchmark-type}), to which
the available datasets can be assigned. It can be seen from Table \ref{table-benchmarks} that
there exist at least 123 datasets for ground states of CS-MG\CSMGGS, OS-MG\OSMGGS, and CS/OS-TM\CSOSTMGS
 (including lanthanides\cite{Ln54,LnIP4,LnSLnSe} and actinides\cite{An66,AnIPAngela2022}).
Such datasets were mainly used to calibrate SR methods\GSSRuse, especially density functional theory (DFT)\GSDFTuse, although
some of them were also used to assess the performance of MR methods \cite{LnIP4,LnSLnSe,GS-DBH24-MR2015,TMdissocEneMR2016,DDB_110,DDB_111,DDB_154,G2-HBCI,FCI21,FCIQMC-diatom}.
In parallel, there exist 73 datasets oriented to excited states of CS-MG\CSMGES, OS-MG\OSMGES, CS-TM\CSTMES, and OS-TM\OSTMES,
aiming to calibrate both SR\ESSR and MR\ESMR methods.

{
\scriptsize
\begin{center}
\begin{threeparttable}[]
\caption{Up-to-date datasets\tnote{*}}
\label{table-benchmarks}
\begin{tabular}{lllllll}
\toprule[0.5pt]	
%\rowcolor[gray]{0.8}\multicolumn{2}{c}{\textbf{Ground State}} \\
\multicolumn{2}{c}{\textbf{Ground State}} \\
\toprule[0.5pt]	
CS-MG  & Refs. \CSMGGSnum \\
OS-MG  & Refs. \OSMGGSnum \\
CS-TM  & Refs. \CSTMGSnum \\
OS-TM  & Refs. \OSTMGSnum \\
\toprule[0.5pt]	
%\rowcolor[gray]{0.8}\multicolumn{2}{c}{\textbf{Excited State }} \\
\multicolumn{2}{c}{\textbf{Excited State }}\\
\toprule[0.5pt]	
CS-MG  & Refs. \CSMGESnum \\
OS-MG  & Refs. \OSMGESnum \\
CS-TM  & Refs. \CSTMESnum \\
OS-TM  & Refs. \OSTMESnum \\
\bottomrule[0.5pt]
\end{tabular}
\begin{tablenotes}
\item * CS: closed-shell; OS: open-shell; MG: main group; TM: transition metal.
\end{tablenotes}
\end{threeparttable}
\end{center}
}

As for the reference data, the very first point to be realized lies in that experimental measurements often cannot be taken as they stand.
For instance, experimentally measured 0-0 transition energies cannot directly be compared to theoretically calculated vertical excitation energies (VEE),
which are not observable. Instead, geometric and vibrational effects must be added to theoretical VEEs or subtracted from experimental 0-0 transition energies
\cite{grimme2004calculation, send_adiabatic, benchmarkCC_DFT13, 40dyes2012Jacquemin, 40large2004Grimme,  loos0-0_19JCTC, loos0-0_18, loos0-0_19CPC, V_0_Adiabatic}. The lack of sufficient experimental data necessitates the use of some high-level methods as the reference for low-level methods.
However, this is not always reliable. It is often the case that
the chosen high-level method is itself not sufficiently accurate. For instance,
the near-exact iCIPT2/TZVP (iterative configuration interaction with selection and second-order perturbation\cite{iCIPT2,iCIPT2New})
calculations\cite{benchmarkSDS} of the singlet VEEs of the dataset\cite{Thiel08}
overturned the recommendation of using the CASPT2/TZVP rather than the CC3/TZVP values as
the theoretical best estimates (TBE),\cite{Thiel08} which was based simply on the extent of MR characters.

Close inspections of the up-to-date datasets (see Table \ref{table-benchmarks}) reveal that there exists only one dataset\cite{benchmarkSDS} for calibrating MR methods
for the excited states of open-shell organic systems. As such, more comprehensive datasets are highly desired.
As a first try towards this goal, we start with the QUEST\#4 dataset\cite{QUEST4}, which contains 24 organic radicals (see Fig. S1 in the Supporting Information).
Since only 51 doublet excited states were reported therein, an immediate
extension is to cover more excited states, both doublet and quartet. Such an extended dataset, covering 110 doublet and 39 quartet states
for the 24 radicals,
is to be dubbed QUEST\#4X. The VEEs calculated by iCIPT2 will be employed to calibrate MR methods
[SS/MS-NEVPT2 (single/multi-state second-order
n-electron valence state perturbation theory)\cite{NEVPT2,MSNEVPT2}, SDSPT2 (static-dynamic-static (SDS) second-order
perturbation theory)\cite{SDS,SDSPT2}, SDSCI\cite{SDS}, and ic-MRCISD
(internally contracted multireference configuration interaction with singles and doubles)\cite{XianCI2018}], as well as
the spin-adapted open-shell TD-DFT\cite{SA-TDDFT1,SA-TDDFT2,XTDDFT}.
The SDSCI, SDSPT2, and iCIPT2 methods are first recapitulated in Sec. \ref{Methods},
which is followed by computational details in Sec. \ref{COMPUTATIONAL}.
The results are analyzed in Sec. \ref{Results}. The paper is closed with concluding remarks in Sec. \ref{Conclusion}.

\section{Computational Methods}\label{Methods}
The (restricted) static-dynamic-static (SDS) family of methods for strongly correlated systems of electrons has been presented 
at length before\cite{benchmarkSDS,LiuWIRES2023}.
Therefore, it is only necessary to explain here their internal relations. In the first place, no matter how many electrons and how many orbitals
are to be correlated, SDSCI\cite{SDS} always constructs and diagonalizes a $3N_P$-by-$3N_P$ matrix for $N_P$ states simultaneously.
Therefore, SDSCI is a minimal MRCI resulting from the combination of the good of perturbation theory, intermediate Hamiltonian, and configuration interaction.
Taking the $N_P$ eigenvectors of SDSCI as new references and repeating SDSCI until convergence,
we obtain iCI\cite{iCI}, where each iteration accesses a space that is higher by two ranks than
that of the preceding iteration. That is, up to $2M$-tuple excitations
(relative to the initial reference space) can be accessed if $M$
iterations are carried out. Because of the variational nature, any
minor loss of accuracy stemming from the contractions can be
removed by carrying out some micro-iterations. In other words,
by controlling the numbers of macro- and micro-iterations, iCI
will generate a series of contracted/uncontracted single/multireference CISD$\cdots 2M$, with the resulting energy being
physically meaningful at each level. It has been shown both theoretically
and numerically that iCI can converge monotonically and quickly from above to FCI, even
when starting with a very poor initial guess. As such,
iCI can be interpreted as an exact solver of FCI. Further combined with the selection of configurations
for static correlation and perturbation correction for dynamic correlation, we obtain the near-exact
iCIPT2\cite{iCIPT2,iCIPT2New}. On the other hand, if the $QHQ$ block of the SDSCI matrix (i.e.,
matrix elements within the first-order interacting space) is replaced by $QH_0Q$, we obtain
SDSPT2\cite{SDS,SDSPT2}, a CI-like MRPT2 that treats single and multiple states in the same way and
is particularly advantageous when a number of states are nearly degenerate (because of
the sufficient relaxation of the reference coefficients)\cite{SDSPT2}. In short, if the
SDSCI calculation is to be performed, we would obtain SDSPT2 results for free.
The latter further yields, e.g., MS-NEVPT2 results for free, because all matrix elements
required by MS-NEVPT2\cite{MSNEVPT2} are already available. Moreover, SDSCI can be taken
as a start of ic-MRCISD, so as to facilitate the convergence of the latter.
As a matter of fact, the increments of the ic-MRCISD iterations clearly indicate the accuracy of SDSCI.
Given so many good features, neither SDSCI nor SDSPT2 is size consistent. However,
the errors can readily be cured\cite{benchmarkSDS} by using the Pople correction\cite{pople1977}.
This is also the case for ic-MRCISD\cite{XianCI2018}. Therefore, the SDSPT2, SDSCI, and ic-MRCISD results
reported here all refer to those with the Pople corrections.
\section{Computational Details}\label{COMPUTATIONAL}
The 24 radicals and their geometries in the QUEST\#4 dataset\cite{QUEST4} were held unchanged in the
iCIPT2\cite{iCIPT2,iCIPT2New}, ic-MRCISD\cite{XianCI2018}, SDSCI\cite{SDS,SDSPT2}, SDSPT2\cite{SDSPT2}, and (MS-)NEVPT2\cite{NEVPT2,MSNEVPT2}
calculations carried out with the BDF program package\cite{BDF1,BDF2,BDF3,BDFECC,BDFrev2020} under the highest Abelian group symmetries.
Since the aug-cc-pVTZ (AVTZ) basis sets\cite{AVTZ1989,AVTZ1992,AVTZ1993} are good enough for most of the excitation energies
(with the mean deviation (MD) only of 0.02 eV from the complete basis set (CBS) limit)\cite{QUEST4},
they were also used here. SA-CASSCF (state-averaged complete active space self-consistent field) calculations
with equal weights for all states were first carried out (see Sec.\ref{SecActSpace} for the various active spaces).
The Dyall Hamiltonian\cite{Dyall-H} was then diagonalized within the chosen active space and used as the active part of 
the zeroth-order Hamiltonian $H_0$ in both SDSPT2 and (MS-)NEVPT2.
The inactive part ($H_{in}$) of $H_0$ is composed of orbital energies for the doubly and zero occupied orbitals. They were
obtained by diagonalizing the generalized Fock matrix (constructed
with the state-averaged one-particle density matrix (1RDM)) within the doubly and zero occupied subspaces separately.
However, such a choice of $H_{in}$ is problematic for states dominated by Rydberg characters (more than 70\%)
 when they were averaged equally with valence states in the SA-CASSCF calculations.
The reason is very simple: Rydberg states stem from configurations very different from those of valence states.
As such, it should be better to use state-specific orbital energies, that is, the doubly and zero occupied orbital energies
are to be determined by the Fock matrix that is constructed with the 1RDM of each Rydberg state itself.
Conceptually, the resulting state-specific orbitals should also be used when constructing the SDSPT2/MS-NEVPT2 effective Hamiltonian matrix.
However, this is not only expensive due to additional integral transformations but also means that the off-diagonal Hamiltonian matrix elements would
involve two sets of non-orthogonal orbitals. Therefore, SA-CASSCF instead of state-specific orbitals were still used in such `state-dependent Fock'
for $H_{in}$\cite{NEVPT2benchmark}. Although it is essentially indistinguishable
from the `state-averaged Fock' for valence excitations\cite{NEVPT2benchmark} ,
it is a must for pure Rydberg states (see Table S1 in the Supporting Information).
Nevertheless, the `state-averaged Fock' still performs better
for states of heavily mixed valence and Rydberg characters (e.g., the
$2~^{2}B_{2}$, $3~^{2}B_{2}$, and $4~^{2}B_{2}$ states of C$_{3}$H$_{5}$ are roughly 1:1 mixtures of valence and Rydberg excitations).

Since iCIPT2 samples the whole Hilbert space, the initial SA-CASSCF calculation is not really needed, although it does provide a good start for the selection procedure.
Anyway, iCIPT2 can work with natural orbitals (NOs) obtained by diagonalizing the 1RDM for the selected configurations with, e.g., $C_{\mathrm{min}}=7 \times 10^{-5}$.
Here, $C_{\mathrm{min}}$ is the single parameter involved in iCIPT2. That is, all configuration state functions
with coefficients smaller in absolute value than $C_{\mathrm{min}}$ will be pruned away from the variational space determined iteratively 
(for more details, see Ref. \citenum{iCIPT2New}). With the fixed NOs, a series of iCIPT2 calculations can then 
be performed with decreasing $C_{\mathrm{min}}$, so as to approach the FCI limit ($C_{\mathrm{min}}=0$) by linear fit of the iCIPT2 vs PT2 energy plot.
For both the ground and excited states, the extrapolation uncertainties at the 95\% confidence level were kept below 1 mHartree. 
Such uncertainties for the ground and excited states were then summed up as the error bars
(less than 2 mHartree or 0.05 eV) for the extrapolated VEEs. 
The extrapolation distances, i.e., the differences between the largest iCIPT2 calculations (at $C_{\mathrm{min}}=1 \times 10^{-5}$)
 and extrapolated values, are also provided for reference. 

%It turns out that $C_{\mathrm{min}}=(4, 3, 2 , 1) \times 10^{-5}$  should be used for
%C$_{3}$H$_{5}$, CNO, CON, and CH$_{2}$NO$_{2}$, whereas $C_{\mathrm{min}}=(7, 5, 3, 1) \times 10^{-5}$
%are already sufficient for other systems.

%It should also be mentioned that the $1s$ and $1s2s2p$ orbitals for the second- and third-row elements, respectively,
%第二行的Be没冻结
It should also be mentioned that the core orbitals were kept frozen in the correlation calculations.

Finally, calculations with the spin-adapted open-shell TD-DFT\cite{SA-TDDFT1,SA-TDDFT2,XTDDFT}
under the Tamm-Dancoff approximation (denoted as X-TDA), in conjunction with the BHandHLYP functional \cite{becke1988,lee1988,becke1993}
(the simplest yet reliable functional for the low-lying doublet-doublet\cite{LZD1} and doublet-quartet excitations\cite{LZD2} of radicals),
were also performed for comparison.

\section{Results and Discussion}\label{Results}

\subsection{Reference data}\label{ReferenceData}
\begin{figure}
	\centering
	\includegraphics[width=0.6\textwidth]{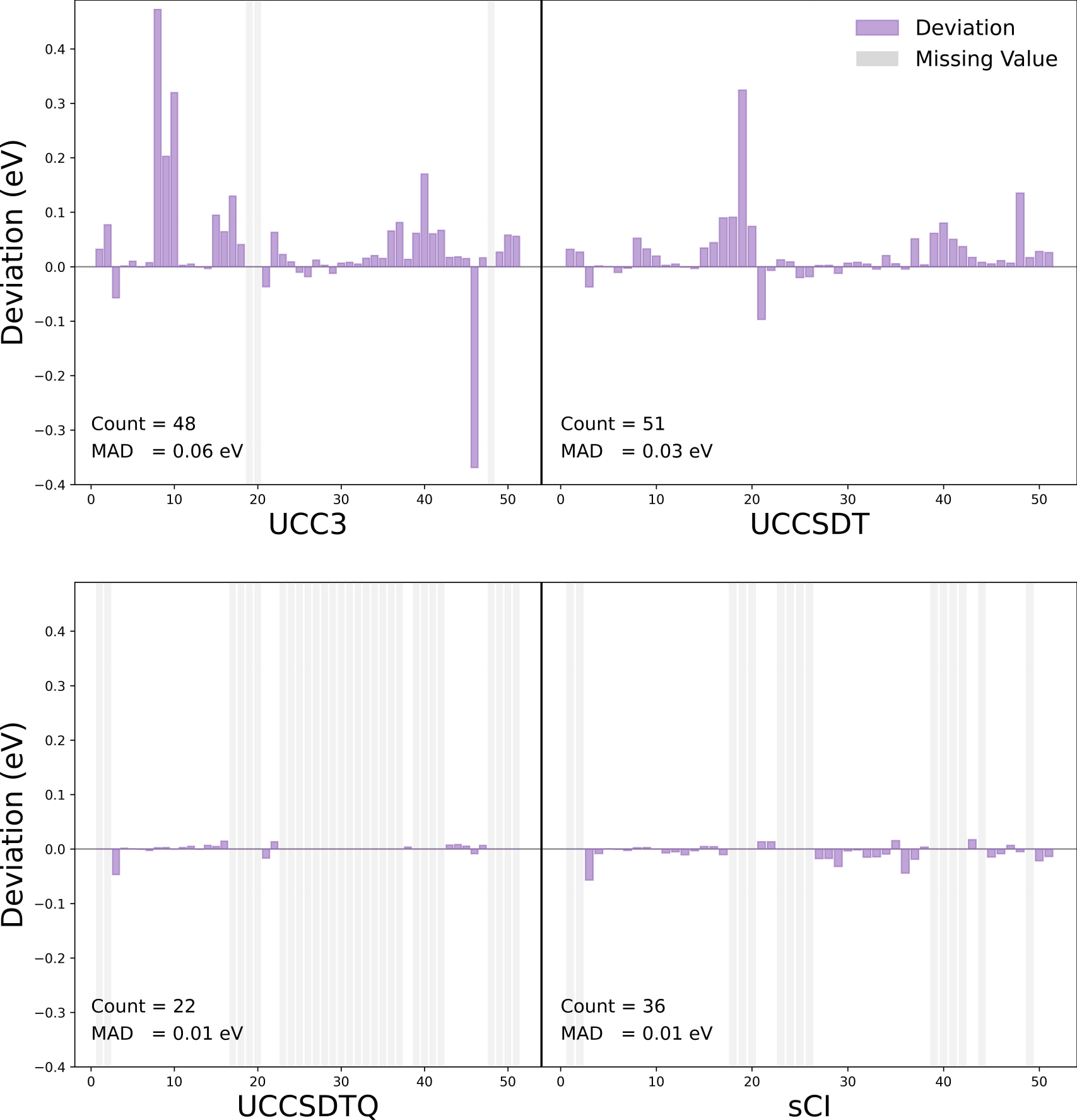}
	\caption{ Deviations (in eV) of UCC3, UCCSDT, UCCSDTQ, and sCI
from iCIPT2 for the 51 doublet states in QUEST\#4 (AVTZ results)\cite{QUEST4}. 
Grey bars represent missing values.}
	\label{fig-QUEST51-ici}
\end{figure}

Ideally, well-resolved experimental data can be taken as the reference. However, for the 24 radicals considered here,
only some experimental 0-0 transitions\cite{C3H5_EXP06, diatomic_EXP, ExptGeom2, CH_EXP, Herzberg1950Molecular, CNO_EXP, F2BO_EXP_CC, X2BY_JCP15, H2PS_EXP, NCO_EXP1, PH2_EXP_1995, C2H3_EXP}, after correcting the geometric and vibrational effects\cite{loos0-0_19JCTC, openshellGordon,CH_EXP}, are available for comparison (see Table \ref{EXPdata}).
Nevertheless, they are enough to verify the accuracy of iCIPT2. It can be seen from Table \ref{EXPdata}
that iCIPT2, in conjunction with the AVTZ basis sets\cite{AVTZ1989,AVTZ1992,AVTZ1993}, is indeed very accurate, except for
the $2~^{2}A_{2}''$ state of CH$_3$ and the $2~^{2}\Sigma^{+}$ state of NO. It turns out that there exist
substantial basis set effects for the two states\cite{QUEST4}. When the AV5Z basis set\cite{AVTZ1989,AVTZ1992,AVTZ1993} is used,
the errors for the two states are reduced from 0.21 and 0.25 eV
to 0.04 and 0.06 eV, respectively. Not surprisingly, the iCIPT2 results are very close to those by selected CI (sCI)\cite{QUEST4} 
%remove CIPSI here
%(configuration interaction using a perturbative selection made iteratively)\cite{CIPSI2019}, 
since both methods are nearly exact. The deviations of the methods (UCC3, UCCSDT, UCCSDTQ, and sCI) adopted in Ref. \citenum{QUEST4} from iCIPT2
are further plotted in Fig. \ref{fig-QUEST51-ici} for the 51 doublets reported therein. All these
pinpoint that the iCIPT2/AVTZ VEEs for the 110 doublets and 39 quartets of the 24 radicals can be
taken as a solid reference for calibrating other methods with the same basis sets.

%Note that the TBE/CBS excitation energy of $2~^{2}A_{2}''$ state of CH$_3$ reported by Loos and coworkers (CIPSI results corrected by CCSDT/AV5Z-AVTZ) is found to be exactly the same as that by iCIPT2/AV5Z.\cite{QUEST4}
%However, for NO radical, the error of TBE/CBS (CIPSI results corrected by CCSDTQ/AVQZ-AVTZ) is larger than that of iCIPT2/AV5Z, indicating that the quadruple-$\zeta$ basis set may not be sufficiently large.

%
%And in the present work, we extend this effort by offering reference data at the same level of accuracy for 105 doublets and 37 quartets using iCIPT2 method.
%Although there is no actual need to benchmark the nearly exact iCIPT2 methods, we decided to validate the accuracy of iCIPT2 through comparisons with both theoretical and experimental data.
%As shown in Fig. \ref{fig-QUEST51-ici}, the CIPSI/AVTZ results in QUEST\#4\cite{QUEST4} cover only 36 states.
%In contrast, our study provides iCIPT2/AVTZ results, without basis set correction, for all 51 doublets in QUEST\#4.
%As expected, CIPSI and iCIPT2 show excellent consistency, with MAD of 0.01 eV.
%This high level of agreement is also observed between UCCSDTQ and iCIPT2.
%The mutual consistency among these high-level calculations demonstrates the reliability of iCIPT2 from a theoretical perspective.
%%Additionly, the accuracy of UCC3 is very close to that of UCCSDT, with MAD of 0.06 eV and 0.03 eV, respectively.

\newpage

{											
	\scriptsize
	\begin{center}
	\begin{threeparttable}[]
			\caption{Experimental and theoretical vertical excitation energies (VEEs in eV) of 26 doublet states}
\label{EXPdata}
\begin{tabular}{lccccccc}
\toprule[0.5pt]	
\textbf{Radical}&\textbf{State}&\textbf{Expt.}&\textbf{iCIPT2/AVTZ}\tnote{a} &\textbf{95\% interval}\tnote{b} &\textbf{CCSDT/AVTZ}\tnote{c} & \textbf{sCI/AVTZ}\tnote{a,d}&\textbf{TBE/CBS}\tnote{e} \\
\toprule[0.5pt]
C$_{3}$H$_{5}$  & $1~^{2}B_{1}$ & 3.41    \tnote{f}   & 3.40   $\pm$ 0.02  & $\pm$ 0.03    & 3.43 &  -         & - \\
BeF & $2~^{2}\Sigma^{+}$ & 6.16   \tnote{g}  & 6.27   $\pm$ 0.00\tnote{h}  & $\pm$ 0.01    & 6.23 & 6.21   $\pm$ 0.02  & -   \\
& $1~^{2}\Pi$ & 4.14   \tnote{g}  & 4.15   $\pm$ 0.00  & $\pm$ 0.00    & 4.15 & 4.14   $\pm$ 0.01  & 4.13  \\ 
BeH & $1~^{2}\Pi$ & 2.48   \tnote{g}  & 2.49   $\pm$ 0.00  & $\pm$ 0.00    & 2.49 & 2.49   $\pm$ 0.00  & 2.48  \\ 
& $2~^{2}\Pi$ & 6.32  \tnote{g}  & 6.46   $\pm$ 0.00  & $\pm$ 0.00    & 6.45 & 6.46   $\pm$ 0.00  & 6.46  \\ 
CH & $1~^{2}\Sigma^{+}$ & 3.94   \tnote{g}  & 3.98   $\pm$ 0.00  & $\pm$ 0.00    & 4.03 & 3.98   $\pm$ 0.00  & 3.96  \\ 
& $1~^{2}\Delta$ & 2.88   \tnote{g}  & 2.91   $\pm$ 0.00  & $\pm$ 0.00    & 2.94 & 2.91   $\pm$ 0.00  & 2.90  \\ 
& $1~^{2}\Sigma^{-}$ & 3.26   \tnote{i}  & 3.29   $\pm$ 0.00  & $\pm$ 0.00    & 3.31 & 3.29   $\pm$ 0.00  & 3.28  \\ 
CH$_{3}$ & $1~^{2}A_{1}'$ & 5.73   \tnote{g}  & 5.86   $\pm$ 0.00 (5.88 $\pm$ 0.00) & $\pm$ 0.00   ( $\pm$ 0.01) & 5.86 & 5.85   $\pm$ 0.01  & 5.88  \\ 
%&  &   & (5.88 $\pm$ 0.00)\tnote{a}    & $\pm$ 0.01    & &     &    \\
& $2~^{2}A_{2}''$ & 7.44   \tnote{g}  & 7.65   $\pm$ 0.00 (7.48$\pm$ 0.00)  & $\pm$ 0.00 ( $\pm$ 0.01)   & 7.65 & 7.65   $\pm$ 0.01  & 7.48  \\ 
%&  &   & (7.48$\pm$ 0.00)\tnote{a}     & $\pm$ 0.01    & &     &    \\
CN & $2~^{2}\Sigma^{+}$ & 3.22   \tnote{g}  & 3.22   $\pm$ 0.00  & $\pm$ 0.00    & 3.25 & 3.22   $\pm$ 0.00  & 3.21  \\ 
& $1~^{2}\Pi$ & 1.32   \tnote{g}  & 1.34   $\pm$ 0.00  & $\pm$ 0.00    & 1.38 & 1.34   $\pm$ 0.01  & 1.33  \\ 
CNO & $1~^{2}\Sigma^{+}$ & 1.55   \tnote{j}  & 1.62   $\pm$ 0.01  & $\pm$ 0.00    & 1.71 & 1.61   $\pm$ 0.01  & 1.61  \\ 
CO$^{+}$ & $2~^{2}\Sigma^{+}$ & 5.82   \tnote{g}  & 5.80   $\pm$ 0.00  & $\pm$ 0.00    & 5.70 & 5.81   $\pm$ 0.00  & 5.80  \\ 
& $1~^{2}\Pi$ & 3.26   \tnote{g}  & 3.27   $\pm$ 0.00  & $\pm$ 0.00    & 3.26 & 3.28   $\pm$ 0.00  & 3.26  \\ 
F$_{2}$BO & $1~^{2}A_{1}$ & 2.77   \tnote{k}  & 2.77   $\pm$ 0.01  & $\pm$ 0.04    & 2.78 &  -      & - \\
& $1~^{2}B_{1}$ & 0.70   \tnote{k}  & 0.70   $\pm$ 0.04  & $\pm$ 0.03    & 0.71 &  -      & - \\
F$_{2}$BS & $1~^{2}A_{1}$ & 2.91   \tnote{l}  & 2.95   $\pm$ 0.02  & $\pm$ 0.05    & 2.93 &  -      & - \\
& $1~^{2}B_{1}$ & 0.46   \tnote{l}  & 0.50   $\pm$ 0.00  & $\pm$ 0.04    & 0.48 &  -      & - \\
H$_{2}$PS & $2~^{2}A'$ & 2.77   \tnote{m}  & 2.73   $\pm$ 0.01  & $\pm$ 0.01    & 2.75 & 2.72   $\pm$ 0.02  & 2.71  \\ 
NCO & $1~^{2}\Sigma^{+}$ & 2.80  \tnote{n} & 2.87   $\pm$ 0.01  & $\pm$ 0.03    & 2.87 & 2.83   $\pm$ 0.05  & 2.89  \\ 
NO & $1~^{2}\Sigma^{+}$ & 5.92   \tnote{g}  & 6.11   $\pm$ 0.01 (6.05$\pm$ 0.01)  & $\pm$ 0.01 ($\pm$ 0.04)   & 6.13 & 6.13   $\pm$ 0.02  & 6.12  \\ 
%&  &   & (6.05$\pm$ 0.01)\tnote{a}     & $\pm$ 0.04    & &     &    \\
& $2~^{2}\Sigma^{+}$ & 7.03   \tnote{g}  & 7.28   $\pm$ 0.01 (7.09$\pm$ 0.01) & $\pm$ 0.00 ( $\pm$ 0.04)   & 7.29 & -        & [7.21]\\ 
%&  &   & (7.09$\pm$ 0.01)\tnote{a}     & $\pm$ 0.04    & &     &    \\
OH & $1~^{2}\Sigma^{+}$ & 4.09   \tnote{g}  & 4.11   $\pm$ 0.00  & $\pm$ 0.00    & 4.12 & 4.10   $\pm$ 0.01  & 4.09  \\ 
PH$_{2}$ & $1~^{2}A_{1}$ & 2.74  \tnote{o} & 2.76   $\pm$ 0.00  & $\pm$ 0.00    & 2.77 & 2.77   $\pm$ 0.00  & 2.76  \\ 
C$_{2}$H$_{3}$ & $1~^{2}A''$ & 3.22  \tnote{p} & 3.28   $\pm$ 0.00  & $\pm$ 0.01    & 3.31 & 3.26   $\pm$ 0.02  & - \\

%Count\tnote{n} &&& 26 &&& 26   &  20  & 19\\
MD\tnote{q}  &&& 0.05 && 0.06 & 0.05 &  0.04  \\
MAD\tnote{q} &&& 0.06 && 0.06 & 0.06  & 0.06  \\
SD\tnote{q}  &&& 0.09 && 0.10 & 0.08  & 0.08  \\
MAX\tnote{q} &&& 0.25 && 0.26 & 0.21  & 0.18  \\
\bottomrule[0.5pt]
\end{tabular}
\begin{tablenotes}
\item[a]$\pm x$ means extrapolation distance $|x|$. In parentheses are the AV5Z results.
\textsuperscript{b}Sum of the 95\% confidence intervals of the ground and excited states. In parentheses are the AV5Z results.
\textsuperscript{c}Ref. \citenum{QUEST4}.
\textsuperscript{d}Ref. \citenum{QUEST4}.
\textsuperscript{e}Ref. \citenum{QUEST4}. sCI results calculated or corrected to at least AVQZ level. In brackets is the CCSDTQ/AVQZ result.
\textsuperscript{f}Refs. \citenum{loos0-0_19JCTC,C3H5_EXP06}.
\textsuperscript{g}Refs. \citenum{openshellGordon,diatomic_EXP, ExptGeom2}.
\textsuperscript{h}6.22 $\pm$ 0.00 eV with the AVTZ from Basis Set Exchange\cite{BSE}. %The AVTZ basis set of Be varies across different program packages.
\textsuperscript{i}Refs. \citenum{CH_EXP,Herzberg1950Molecular}.
\textsuperscript{j}Refs. \citenum{loos0-0_19JCTC,CNO_EXP}.
\textsuperscript{k}Refs. \citenum{loos0-0_19JCTC,F2BO_EXP_CC}.
\textsuperscript{l}Refs. \citenum{loos0-0_19JCTC,X2BY_JCP15}.
\textsuperscript{m}Refs. \citenum{loos0-0_19JCTC,H2PS_EXP}.
\textsuperscript{n}Refs. \citenum{loos0-0_19JCTC,NCO_EXP1}.
\textsuperscript{o}Refs. \citenum{loos0-0_19JCTC,PH2_EXP_1995}.
\textsuperscript{p}Refs. \citenum{loos0-0_19JCTC,C2H3_EXP}.
\textsuperscript{q}MD: mean deviation; MAD: mean absolute deviation; SD: standard deviation; MAX: maximum deviation.
Unavailable data were excluded from the analysis.
\end{tablenotes}			
\end{threeparttable}
\end{center}
}

\subsection{QUEST\#4X}
Some criteria should first be established for adding additional excited states to the QUEST\#4 dataset\cite{QUEST4}.
Since the highest VEE reported therein is 8.02 eV ($^{2}\Sigma^{-}$ of OH), only
those singly excited doublet and quartet states with VEEs up to 8.0 eV are to be considered here.
To identify such states, iCIPT2/AVTZ calculations were performed on the low-lying excited states of the 24 radicals for each irreducible representation (irrep).
%A special situation occurred to C$_3$H$_5$, which has eight doublet and two quartet states in addition to
%the $1~^{2}A_{1}$ and $1~^{2}B_{2}$ states already in QUEST\#4.
Eventually, 59 doublet states (including 17 Rydberg states) and 39 quartet states (including 6 Rydberg states)
were added to QUEST\#4 (which has 51 doublet states, including 11 Rydberg states), leading to QUEST\#4X.
Although the AVTZ basis sets may not be sufficient for some of the Rydberg states,
the iCIPT2/AVTZ results can certainly be taken as the reference for calibrating other methods with the same basis sets.

\subsection{Active space}
\label{SecActSpace}
Unlike iCIPT2, the NEVPT2, SDSPT2, SDSCI, and ic-MRCISD calculations all require a proper choice of active spaces.
The occupation numbers (ON) of the natural orbitals (NO) resulting from the iCIPT2 calculations were taken here as guidance. 
Those NOs with ONs ranging from 1.98 to 0.02 were carefully analyzed as candidates for active orbitals.
However, to make the MR calculations feasible, the number of active orbitals was kept below 15 (by deleting those virtual orbitals
of the smallest ONs). 
The resulting active spaces for the 24 radicals are documented in Table \ref{table-ActiveSpaces}.
The SA-CASSCF calculations can be initiated and supervised by the iCAS approach (imposed automatic selection of complete active spaces)\cite{iCAS}, where once the valence atomic orbitals (see Table S2 in the Supporting Information) are prepared, all remaining steps are automated. 
In particular, the subspace matching algorithm therein
preserves the characters of the orbitals (doubly occupied, active, or virtual) throughout the self-consistent iterations.
The CASSCF energies are documented in Table S3 in the Supporting Information.
%\begin{enumerate}[(1)]
%\item For radicals containing one non-hydrogen (non-H) atom, the active spaces
%comprise all valence and next shell orbitals $nsnp(n+1)s(n+1)p$ of non-H atoms, alongside the hydrogen $1s$.
%\item For radicals containing two non-H atoms, the active spaces encompass all valence orbitals $nsnp$ and the Rydberg orbitals involved in the excitations.
%\item For radicals consisting of three non-H atoms and no hydrogen atom, all valence orbitals of non-H atoms were included.
%\item For the remaining three radicals with 4 non-H atoms, the VAOs are given in Table \ref{VAOs}.
%\end{enumerate}

{		
\scriptsize
\begin{center}		
\begin{threeparttable}[]
\caption{Active spaces and electronic states by SA-CASSCF}
\label{table-ActiveSpaces}
\begin{tabular}{lccccc}
\hline

\textbf{Radical}	&\textbf{	Active Space}	&	\textbf{Symmetry}	&	\textbf{Irrep of Active MO}\tnote{a}	&	\multicolumn{2}{c}{\textbf{Electronic State}\tnote{b}}\\
	&	         	&	           	&  	                 	&	\textbf{Doublet}	&	\textbf{Quartet} 	\\
%  	&		         &	          	&$C_{s}$ ($a',a''$)  		&\multicolumn{2}{c}{$C_{s}$ ($A',A''$)}\\
%  	&	           	&	           	&$C_{2v}$ ($a_1,a_2,b_1,b_2$) &	\multicolumn{2}{c}{$C_{2v}$ ($A_1,A_2,B_1,B_2$)}\\
 \hline
C$_{3}$H$_{5}$	&	(3e,9o)	&	$C_{2v}$	&	(2,2,1,4)	&	(2,1,1,4)	&	(0,1,1,0)	\\
BeF	&	(9e,10o)	&	$C_{2v}$	&	(6,0,2,2)	&	(2,0,1,1)	&	-	\\
BeH	&	(3e,11o)	&	$C_{2v}$	&	(5,0,3,3)	&	(3,0,4,4)	&	(0,0,1,1)	\\
BH$_{2}$	&	(5e,10o)	&	$C_{2v}$	&	(5,0,3,2)	&	(3,2,2,2)	&	(0,1,0,0)	\\
CH	&	(5e,9o)	&	$C_{2v}$	&	(5,0,2,2)	&	(3,2,1,1)	&	(0,1,1,1)	\\
CH$_{3}$	&	(7e,11o)	&	$C_{2v}$	&	(6,0,3,2)	&	(3,0,2,2)	&	-	\\
CN	&	(9e,8o)	&	$C_{2v}$	&	(4,0,2,2)	&	(2,0,2,2)	&	(2,2,0,0)	\\
CNO	&	(15e,12o)	&	$C_{2v}$	&	(6,0,3,3)	&	(3,2,3,3)	&	(0,0,1,1)	\\
CON	&	(15e,12o)	&	$C_{2v}$	&	(6,0,3,3)	&	(1,1,2,2)	&	(0,0,1,1)	\\
CO$^{+}$	&	(9e,8o)	&	$C_{2v}$	&	(4,0,2,2)	&	(2,0,1,1)	&	(1,0,0,0)	\\
F$_{2}$BO	&	(19e,14o)	&	$C_{2v}$	&	(6,1,4,3)	&	(2,1,2,1)	&	-	\\
F$_{2}$BS	&	(19e,14o)	&	$C_{2v}$	&	(6,1,4,3)	&	(1,0,2,1)	&	(0,0,1,0)	\\
H$_{2}$BO	&	(11e,10o)	&	$C_{2v}$	&	(5,0,3,2)	&	(2,0,3,2)	&	(0,0,1,0)	\\
HCO	&	(11e,10o)	&	$C_{s}$	&	(8,2)	&	(3,3)	&	(1,1)	\\
HOC	&	(11e,10o)	&	$C_{s}$	&	(8,2)	&	(3,3)	&	(1,1)	\\
H$_{2}$PO	&	(13e,10o)	&	$C_{s}$	&	(7,3)	&	(4,3)	&	(2,1)	\\
H$_{2}$PS	&	(13e,10o)	&	$C_{s}$	&	(7,3)	&	(3,2)	&	(3,3)	\\
NCO	&	(15e,12o)	&	$C_{2v}$	&	(6,0,3,3)	&	(2,0,3,3)	&	(0,0,1,1)	\\
NH$_{2}$	&	(7e,10o)	&	$C_{2v}$	&	(5,0,3,2)	&	(2,0,1,2)	&	(0,0,0,1)	\\
CH$_{2}$NO$_{2}$	&	(13e,10o)	&	$C_{2v}$	&	(3,1,3,3)	&	(2,2,2,2)	&	(1,1,1,0)	\\
NO	&	(11e,10o)	&	$C_{2v}$	&	(6,0,2,2)	&	(2,0,2,2)	&	(0,1,1,1)	\\
OH	&	(7e,9o)	&	$C_{2v}$	&	(5,0,2,2)	&	(1,1,1,1)	&	(0,1,0,0)	\\
PH$_{2}$	&	(7e,10o)	&	$C_{2v}$	&	(5,0,3,2)	&	(2,1,2,2)	&	(0,1,0,1)	\\
 C$_{2}$H$_{3}$	&	(11e,12o)	&	$C_{s}$	&	(10,2)	&	(3,3)	&	(1,1)	\\
\hline
\end{tabular}
\begin{tablenotes}
	\item[a]$C_{s}$: ($a',a''$); $C_{2v}$: ($a_1,a_2,b_2,b_1$). \textsuperscript{b}$C_{s}$: ($A',A''$); $C_{2v}$: ($A_1,A_2,B_2,B_1$).
%	\item[b]$C_{s}$: ($A',A''$); $C_{2v}$: ($A_1,A_2,B_1,B_2$).
\end{tablenotes}
\end{threeparttable}
\end{center}
}

\subsection{Vertical Excitation Energies}
\begin{figure}
	\centering
	\includegraphics[width=1\textwidth]{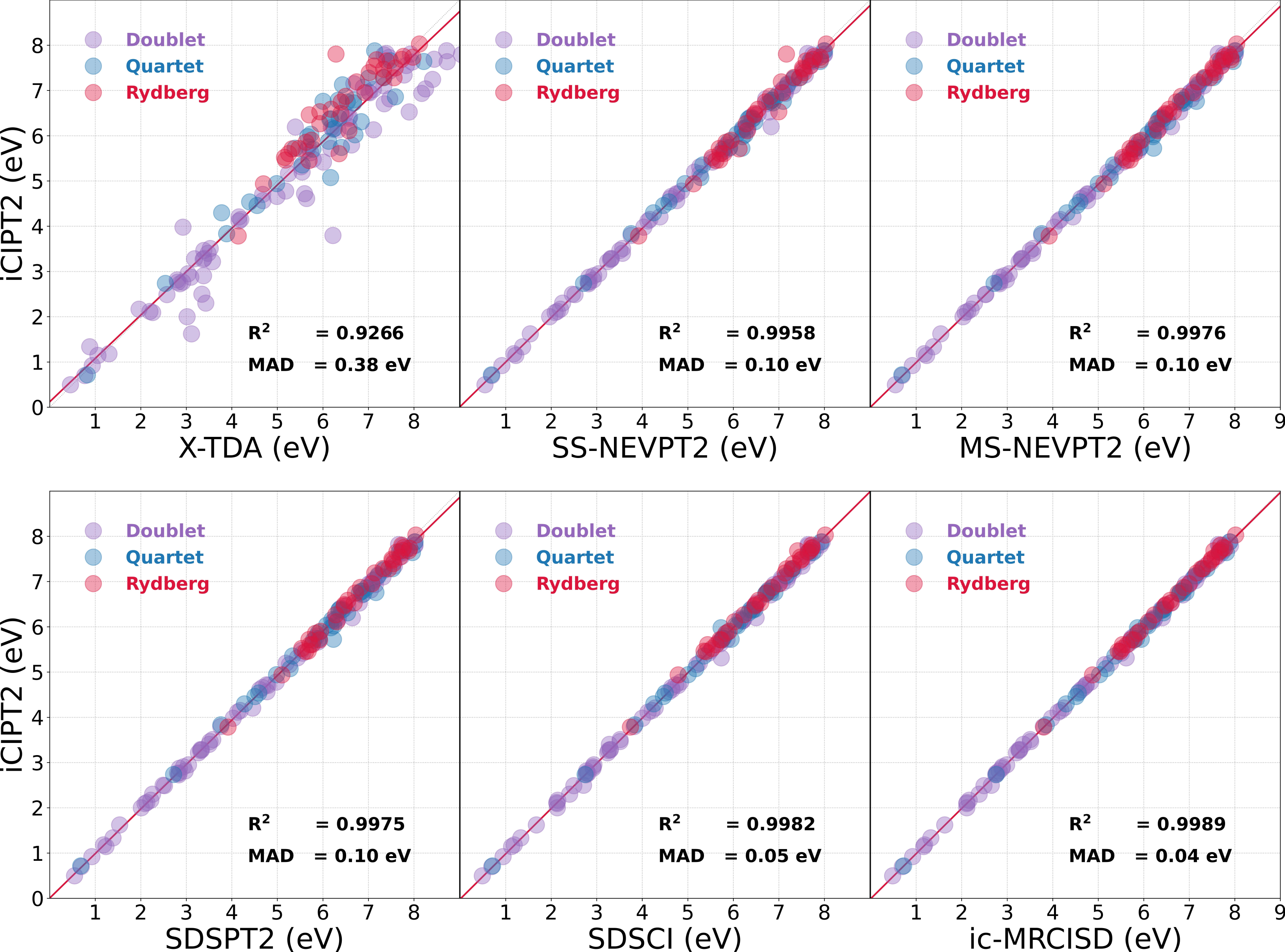}
	\caption{Scatter plots for the iCIPT2 and X-TDA/NEVPT2/SDSPT2/SDSCI/ic-MRCISD VEEs for the QUEST\#4X dataset
		(MAD: mean absolute deviation)}
	\label{fig_radical_line}
\end{figure}

The SS-NEVPT2, MS-NEVPT2, SDSPT2, SDSCI, and ic-MRCISD calculations on the VEEs of the 110 doublet and 39 quartet states
of the 24 radicals in QUEST\#4\cite{QUEST4} were performed with the above setup of active spaces.
The so-calculated VEEs are documented in Table \ref{table-ene}.
Apart from the overall error analysis shown in Table \ref{table-wrtici}, the internal consistency of the methods is also analyzed
in Table \ref{table-devrange}.
Scatter plots for the iCIPT2 and X-TDA/SS-NEVPT2/MS-NEVPT2/SDSPT2/SDSCI/ic-MRCISD VEEs are further depicted in Fig. \ref{fig_radical_line}.

It can be seen from Fig. \ref{fig_radical_line} that
the three variants of MRPT2 (i.e., SS-NEVPT2, MS-NEVPT2, and SDSPT2) have a strong linear correlation with iCIPT2,
with R$^2$ larger than 0.995, which is accompanied by a mean absolute deviation (MAD) of ca. 0.1 eV.
As can be seen from Table \ref{table-devrange}, the VEEs calculated by the three MRPT2 are closely aligned
for more than 80\% of the excited states, with the differences in between being less than 0.05 eV.
However, SS-NEVPT2 fails occasionally. For instance,
it fails to reproduce the ordering of the $2~^{2}A''$ (6.83 eV) and $3~^{2}A''$ (6.71 eV) states of H$_{2}$PO.
Despite some overestimates, both MS-NEVPT2 and SDSPT2 reproduce the correct ordering for the two states
(MS-NEVPT2: 6.61 and 7.04 eV; SDSPT2: 6.64 and 7.05 eV), as compared to the iCIPT2 results (6.20 and 6.81 eV).
A similar situation occurs also  to $1~^{2}E'$ ($2~^{2}A_{1}$ and $1~^{2}B_{2}$  under $C_{2v}$) and $2~^{2}E'$ ($3~^{2}A_{1}$ and $2~^{2}B_{2}$ under $C_{2v}$) states of CH$_3$. These represent examples where dynamic correlation revises significantly the coefficients of the reference functions, as shown in Table \ref{table-H2PO}.
Here, it deserves to be emphasized that, although MS-NEVPT2 and SDSPT2 are
hardly distinguishable for the systems considered here (see Table \ref{table-devrange}), SDSPT2 does outperform MS-NEVPT2
for situations with multiple near-degenerate states\cite{SDSPT2}.

Both SDSCI and ic-MRCISD have a perfect linear correlation with iCIPT2, with R$^2$ being ca. 0.999, which is accompanied by
a MAD of ca. 0.05 eV. SDSCI agrees with ic-MRCISD within 0.15 eV for all the excited states,
given that the computational cost of SDSCI is merely that of one iteration of ic-MRCISD. More specifically,
SDSCI agrees with ic-MRCISD
 within  0.05, 0.10, and 0.15 eV for 75\%, 95\%, and 100\% of the excited states, respectively (cf. Table \ref{table-devrange}).
In particular, this is not fortuitous but holds also for other systems\cite{benchmarkSDS}.
However, it deserves to be pointed out that both SDSCI and ic-MRCISD perform worse than the three MRPT2 for the
$2~^{2}B_{2}$ state of nitromethyl (CH$_{2}$NO$_{2}$), see Table \ref{table-ene}.

Compared to the above MR methods, X-TDA/BHandHLYP has a weaker linear correlation with iCIPT2, as reflected by a lower R$^2$ of ca. 0.927.
Moreover, the orderings of several states of
BH$_2$, CH$_3$, H$_{2}$PO, H$_{2}$PS, and C$_{2}$H$_{3}$ were not reproduced correctly, as marked in Table \ref{table-ene}.
Nonetheless, the accuracy of X-TDA for doublet and quartet excited states
of radicals is comparable to that of TD-DFT for singlet and triplet excited states of closed-shell systems, as already noticed before\cite{LZD1,LZD2}.

Having discussed the general trends, we now focus on the allyl radical \ce{C3H5}, which has been investigated extensively both 
experimentally\cite{C3H5_EXP06, C3H5_EXP_CHEN, C3H5_EXP_Currie, C3H5_IP_JACS1978, C3H5_IP_JCP2003} and theoretically\cite{C3H5_CASPT2,C3H5_CC_2010,C3H5_MRCI_03JACS,C3H5_MRCI_10JPCA,C3H5_orbs_09JCP}, 
due to its pivotal role in photochemical reactions.%\cite{C3H5_reacition00,C3H5_reacition92,C3H5_reacition99,C3H5_reaction,C3H5_reaction15}.
 It can be seen from Table \ref{table-C3H5-ene} that seven of the nine excited states are Rydberg states. 
Because of this, the very first point is the proper choice of basis sets with diffuse functions\cite{C3H5_CASPT2,C3H5_CC_2010,C3H5_MRCI_03JACS,C3H5_MRCI_10JPCA},
among which the augmentation of AVTZ with molecule-centered functions (mcf)\cite{C3H5_CC_2010} is arguably most sophisticated. 
As can be seen from Table \ref{table-C3H5-ene}, there does exist substantial basis set effects for 
the higher-lying Rydberg states (-0.49 eV for $3~^{2}B_1$ and -1.44 eV for $4~^{2}B_1$), as revealed by the UCCSD calculations\cite{C3H5_CC_2010}.
It can hence be deduced that the substantial differences (0.33 and 1.01 eV) between iCIPT2 (6.53 and 7.80 eV)
and ROCC3 (6.20 and 6.79 eV)\cite{C3H5_CC_2010} for the two states stem from the deficit of the AVTZ basis set employed here. 
However, the difference (-0.17 eV) between iCIPT2 (5.98 eV) and ROCC3 (6.15 eV)\cite{C3H5_CC_2010} for the $1~^4A_2$ state 
cannot simply be ascribed to basis set effect, for ROCC3 suffers from noticeable spin contamination in this case. 
As for the lowest five doublet states, iCIPT2 agrees with ROCC3\cite{C3H5_CC_2010} within 0. 05 eV. 
On the other hand, the present ic-MRCISD results differ from the iCIPT2 ones by less than 0.15 eV for all the states.
In contrast, the previous second-order MRCI\cite{C3H5_MRCI_03JACS} or ic-MRCISD\cite{C3H5_MRCI_10JPCA}
results are lower than the present ic-MRCISD (iCIPT2) ones by up to 0.2 (0.3) eV. This should be
ascribed to the use of a smaller basis set (DZP vs AVTZ) and a smaller active space [CAS(3e,7o) vs CAS(3e,9o)] in those calculations. 

As for the MRPT2 calculations, it is first noted that, although SS-NEVPT2 
predicts correctly the relative ordering of the four $^{2}B_{1}$ states, 
it overestimates the excitation energies of the $2~^{2}B_{1}$ and $3~^{2}B_{1}$ states by 0.41 eV and 0.47 eV, respectively, 
but underestimates that of $4~^{2}B_{1}$  by 0.64 eV. To understand such failures of standard MRPT2,
we analyzed the wavefunctions in detail. As can be seen from Table \ref{table-C3H5}, the four $^{2}B_{1}$ states 
stem primarily from four transitions, two valence and two Rydberg transitions.
For the $1~^{2}B_{1}$ state, which is predominantly a mixture of the two valence transitions, all MR methods 
yield similar results, with deviations from iCIPT2 by no more than 0.15 eV.
In contrast, the wavefunctions of the other three $^{2}B_{1}$ states are more complex.
At the CASSCF level, both $2~^{2}B_{1}$ and $3~^{2}B_{1}$ are dominated by Rydberg characters (>70\%),
 while $4~^{2}B_{1}$ is more akin to a valence state yet mixed with 32\% Rydberg character.
The situation changes dramatically at the SDSCI and ic-MRCISD levels, where these states exhibit a strong mixing of all the four components, 
resulting in comparable Rydberg characters exceeding 50\% for each state. Such revisions of the reference states
cannot be captured by standard MRPT2 like SS-NEVPT2, but can be captured to some extents by their multi-state/quasi-degenerate variants. 
For instance, SDSPT2 reproduces essentially all the features of SDSCI/ic-MRCISD, whereas MS-NEVPT2 
is much less successful, in the sense that its $2~^{2}B_{1}$ state contains only 25\% Rydberg component,
less than half of that of ic-MRCISD. Nevertheless, there is no significant difference 
in the overall accuracy of MS-NEVPT2 and SDSPT2: compared with iCIPT2, MS-NEVPT2 overestimates the excitation energies 
of the $2~^{2}B_{1}$ to $4~^{2}B_{1}$ states by 0.07, 0.12, and 0.07 eV, respectively, 
while SDSPT2 exhibits deviations of 0.22, 0.15, and -0.07 eV, respectively. 
It appears that MS-CASPT2 also performs well for these states\cite{C3H5_CASPT2}.

{
	\setlength{\tabcolsep}{6pt}
	\scriptsize
	\begin{center}
		\begin{ThreePartTable}
			\begin{longtable}{llcccccccc}
				%\begin{longtable}{llcccccccccc}P{2.2}
				\caption{Deviations (in eV) of X-TDA, SS-NEVPT2, MS-NEVPT2, SDSPT2, SDSCI, and ic-MRCISD from iCIPT2 for the
vertical excitation energies (VEE) of 110 doublet and 39 quartet states}
				\label{table-ene}\\
				\toprule[0.5pt]
				
				\textbf{Radical}	&	\textbf{State}	&	\textbf{X-TDA}&\textbf{SS-NEVPT2}&	\textbf{MS-NEVPT2}&	\textbf{SDSPT2}&	\textbf{SDSCI}&	\textbf{ic-MRCISD}&	\textbf{iCIPT2}\tnote{a}&\textbf{95\% interval}\tnote{b}\\
				\toprule
				\endfirsthead
				
				\caption{continued}\\
				\toprule[0.5pt]
				\textbf{Radical}	&	\textbf{State}	&	\textbf{X-TDA}&\textbf{SS-NEVPT2}&	\textbf{MS-NEVPT2}&	\textbf{SDSPT2}&	\textbf{SDSCI}&	\textbf{ic-MRCISD}&	\textbf{iCIPT2}\tnote{a}&\textbf{95\% interval}\tnote{b}\\
				\toprule
				\endhead
				
				\bottomrule[0.5pt]
				\endfoot
				
				\bottomrule[0.5pt]
				%\insertTableNotes  % tell LaTeX where to insert the contents of "TableNotes"
				\endlastfoot
Allyl 	&	$1~^{2}A_{1}$	&	-0.24 	&	0.19 	&	0.19 	&	0.16 	&	-0.15 	&	-0.06 	&	4.94 	$\pm$	0.04 	& $\pm$	0.03 	\\
(C$_{3}$H$_{5}$)	&	$2~^{2}A_{1}$	&	-0.28 	&	0.24 	&	0.24 	&	0.21 	&	-0.11 	&	-0.03 	&	5.46 	$\pm$	0.06 	& $\pm$	0.03 	\\
&	$1~^{2}B_{2}$	&	-0.36 	&	0.18 	&	0.18 	&	0.16 	&	-0.17 	&	-0.08 	&	5.61 	$\pm$	0.01 	& $\pm$	0.03 	\\
&	$1~^{2}B_{1}$	&	0.07 	&	0.16 	&	0.15 	&	0.10 	&	-0.13 	&	-0.05 	&	3.40 	$\pm$	0.02 	& $\pm$	0.03 	\\
&	$2~^{2}B_{1}$	&	-0.39 	&	0.41 	&	0.07 	&	0.22 	&	0.03 	&	0.08 	&	5.71 	$\pm$	0.05 	& $\pm$	0.02 	\\
&	$3~^{2}B_{1}$	&	-0.60 	&	0.47 	&	0.12 	&	0.15 	&	0.07 	&	0.07 	&	6.53 	$\pm$	0.01 	& $\pm$	0.03 	\\
&	$4~^{2}B_{1}$	&	-1.52 	&	-0.64 	&	0.07 	&	-0.07 	&	-0.08 	&	-0.11 	&	7.80 	$\pm$	0.02 	& $\pm$	0.04 	\\
&	$1~^{4}A_{2}$	&	-0.31 	&	0.22 	&	0.22 	&	0.19 	&	-0.25 	&	-0.12 	&	5.98 	$\pm$	0.02 	& $\pm$	0.02 	\\
&	$1~^{4}B_{2}$	&	-0.50 	&	0.24 	&	0.24 	&	0.21 	&	-0.27 	&	-0.14 	&	7.68 	$\pm$	0.02 	& $\pm$	0.02 	\\
BeF\tnote{c}&	$2~^{2}\Sigma^{+}$	&	-0.33 	&	0.01 	&	0.01 	&	0.01 	&	-0.03 	&	-0.03 	&	6.27 	$\pm$	0.00 	& $\pm$	0.01 	\\
&	$1~^{2}\Pi$	&	0.05 	&	0.02 	&	0.02 	&	0.04 	&	0.06 	&	0.03 	&	4.15 	$\pm$	0.00 	& $\pm$	0.00 	\\
BeH\tnote{c}&	$2~^{2}\Sigma^{+}$	&	-0.37 	&	0.01 	&	0.01 	&	0.01 	&	0.01 	&	0.00 	&	5.52 	$\pm$	0.00 	& $\pm$	0.00 	\\
&	$3~^{2}\Sigma^{+}$	&	0.45 	&	0.19 	&	0.19 	&	0.19 	&	0.14 	&	0.00 	&	5.72 	$\pm$	0.00 	& $\pm$	0.00 	\\
&	$1~^{2}\Pi$	&	0.08 	&	0.03 	&	0.03 	&	0.03 	&	0.01 	&	0.00 	&	2.49 	$\pm$	0.00 	& $\pm$	0.00 	\\
&	$2~^{2}\Pi$	&	-0.76 	&	0.00 	&	0.00 	&	0.01 	&	0.01 	&	0.00 	&	6.46 	$\pm$	0.00 	& $\pm$	0.00 	\\
&	$3~^{2}\Pi$	&	0.44 	&	0.24 	&	0.23 	&	0.23 	&	0.16 	&	0.01 	&	7.34 	$\pm$	0.00 	& $\pm$	0.00 	\\
&	$4~^{2}\Pi$	&	0.12 	&	0.19 	&	0.20 	&	0.21 	&	0.11 	&	-0.03 	&	7.80 	$\pm$	0.00 	& $\pm$	0.00 	\\
&	$1~^{4}\Pi$	&	0.24 	&	0.03 	&	0.03 	&	0.03 	&	-0.01 	&	0.00 	&	5.88 	$\pm$	0.00 	& $\pm$	0.00 	\\
BH$_{2}$	&	$2~^{2}A_{1}$	&	-0.24 	&	-0.03 	&	-0.02 	&	-0.02 	&	-0.01 	&	0.01 	&	5.72 	$\pm$	0.00 	& $\pm$	0.00 	\\
&	$3~^{2}A_{1}$	&	-0.35 	&	-0.04 	&	-0.03 	&	-0.03 	&	-0.02 	&	0.01 	&	6.75 	$\pm$	0.00 	& $\pm$	0.00 	\\
&	$1~^{2}A_{2}$	&	0.17 	&	-0.03 	&	-0.02 	&	-0.02 	&	0.00 	&	0.01 	&	6.41 	$\pm$	0.00 	& $\pm$	0.00 	\\
&	$2~^{2}A_{2}$	&	0.08 	&	0.06 	&	0.06 	&	0.06 	&	0.04 	&	-0.03 	&	6.95 	$\pm$	0.00 	& $\pm$	0.00 	\\
&	$1~^{2}B_{2}$	&	0.19\tnote{d} 	&	0.05 	&	0.05 	&	0.06 	&	0.04 	&	0.00 	&	6.38 	$\pm$	0.00 	& $\pm$	0.00 	\\
&	$2~^{2}B_{2}$	&	-0.40\tnote{d}  	&	-0.04 	&	-0.03 	&	-0.04 	&	-0.03 	&	0.00 	&	6.58 	$\pm$	0.00 	& $\pm$	0.00 	\\
&	$1~^{2}B_{1}$	&	0.12 	&	0.00 	&	0.00 	&	0.00 	&	0.00 	&	0.01 	&	1.18 	$\pm$	0.00 	& $\pm$	0.00 	\\
&	$2~^{2}B_{1}$	&	-0.37 	&	-0.06 	&	-0.05 	&	-0.05 	&	-0.02 	&	0.01 	&	6.87 	$\pm$	0.00 	& $\pm$	0.00 	\\
&	$1~^{4}A_{2}$	&	0.19 	&	-0.03 	&	-0.02 	&	-0.02 	&	0.00 	&	0.01 	&	5.35 	$\pm$	0.00 	& $\pm$	0.00 	\\
CH	&	$1~^{2}\Sigma^{+}$	&	-1.05 	&	0.06 	&	0.05 	&	0.05 	&	0.02 	&	0.00 	&	3.98 	$\pm$	0.00 	& $\pm$	0.00 	\\
&	$2~^{2}\Sigma^{+}$	&	-0.10 	&	-0.02 	&	-0.02 	&	-0.02 	&	-0.01 	&	-0.01 	&	6.49 	$\pm$	0.00 	& $\pm$	0.00 	\\
&	$1~^{2}\Delta$	&	0.47 	&	0.03 	&	0.03 	&	0.03 	&	0.01 	&	-0.01 	&	2.91 	$\pm$	0.00 	& $\pm$	0.00 	\\
&	$1~^{2}\Sigma^{-}$	&	0.10 	&	0.02 	&	0.02 	&	0.03 	&	0.01 	&	-0.01 	&	3.29 	$\pm$	0.00 	& $\pm$	0.00 	\\
&	$1~^{4}\Sigma^{-}$	&	0.11 	&	-0.03 	&	-0.03 	&	-0.04 	&	-0.01 	&	0.01 	&	0.72 	$\pm$	0.00 	& $\pm$	0.00 	\\
&	$1~^{4}\Pi$	&	-0.18 	&	0.03 	&	0.03 	&	0.02 	&	0.00 	&	0.00 	&	7.66 	$\pm$	0.00 	& $\pm$	0.00 	\\
CH$_{3}$	&	$1~^{2}A_{1}'$	&	-0.23 	&	-0.02 	&	-0.01 	&	-0.02 	&	-0.02 	&	0.01 	&	5.86 	$\pm$	0.00 	& $\pm$	0.00 	\\
&	$1~^{2}E'$	&	0.04\tnote{d}  	&	0.17\tnote{d}  	&	0.07 	&	0.09 	&	0.11 	&	0.04 	&	6.96 	$\pm$	0.00 	& $\pm$	0.00 	\\
&	$2~^{2}E'$	&	-0.45\tnote{d}  &	-0.14\tnote{d}  &	-0.01 	&	-0.05 	&	-0.05 	&	-0.03 	&	7.19 	$\pm$	0.00 	& $\pm$	0.00 	\\
&	$2~^{2}A_{2}''$	&	-0.26 	&	-0.06 	&	-0.03 	&	-0.03 	&	-0.01 	&	0.00 	&	7.65 	$\pm$	0.00 	& $\pm$	0.00 	\\
CN	&	$2~^{2}\Sigma^{+}$	&	0.36 	&	0.02 	&	0.04 	&	0.05 	&	0.02 	&	-0.01 	&	3.22 	$\pm$	0.00 	& $\pm$	0.00 	\\
&	$1~^{2}\Pi$	&	-0.46 	&	0.04 	&	0.05 	&	0.05 	&	0.00 	&	-0.01 	&	1.34 	$\pm$	0.00 	& $\pm$	0.00 	\\
&	$2~^{2}\Pi$	&	0.84 	&	0.11 	&	0.12 	&	0.14 	&	0.07 	&	0.01 	&	7.88 	$\pm$	0.00 	& $\pm$	0.00 	\\
&	$1~^{4}\Sigma^{+}$	&	-0.30 	&	0.04 	&	0.05 	&	0.05 	&	0.02 	&	0.02 	&	6.03 	$\pm$	0.00 	& $\pm$	0.00 	\\
&	$1~^{4}\Delta$	&	-0.70 	&	0.07 	&	0.08 	&	0.09 	&	0.03 	&	0.00 	&	7.12 	$\pm$	0.00 	& $\pm$	0.00 	\\
&	$1~^{4}\Sigma^{-}$	&	-0.74 	&	0.13 	&	0.14 	&	0.15 	&	0.05 	&	0.01 	&	7.88 	$\pm$	0.00 	& $\pm$	0.00 	\\
CNO	&	$1~^{2}\Sigma^{+}$	&	1.49 	&	-0.08 	&	-0.08 	&	-0.08 	&	0.05 	&	0.01 	&	1.62 	$\pm$	0.01 	& $\pm$	0.00 	\\
&	$2~^{2}\Sigma^{+}$	&	0.33 	&	0.10 	&	0.11 	&	0.11 	&	0.06 	&	0.08 	&	7.63 	$\pm$	0.03 	& $\pm$	0.01 	\\
&	$1~^{2}\Delta$	&	0.75 	&	0.10 	&	0.11 	&	0.13 	&	0.12 	&	0.07 	&	7.69 	$\pm$	0.02 	& $\pm$	0.01 	\\
&	$1~^{2}\Sigma^{-}$	&	-0.43 	&	-0.19 	&	-0.18 	&	-0.16 	&	-0.18 	&	-0.18 	&	7.82 	$\pm$	0.02 	& $\pm$	0.01 	\\
&	$2~^{2}\Pi$	&	0.31 	&	0.04 	&	0.05 	&	0.05 	&	0.02 	&	0.02 	&	5.48 	$\pm$	0.01 	& $\pm$	0.00 	\\
&	$3~^{2}\Pi$	&	0.00 	&	0.09 	&	0.10 	&	0.11 	&	0.05 	&	0.06 	&	6.18 	$\pm$	0.03 	& $\pm$	0.01 	\\
&	$1~^{4}\Pi$	&	0.08 	&	0.07 	&	0.08 	&	0.09 	&	0.04 	&	0.06 	&	5.70 	$\pm$	0.01 	& $\pm$	0.00 	\\
CON	&	$1~^{2}\Sigma^{+}$	&	2.42 	&	-0.04 	&	-0.04 	&	-0.04 	&	0.03 	&	0.00 	&	3.80 	$\pm$	0.02 	& $\pm$	0.01 	\\
&	$1~^{2}\Sigma^{-}$	&	-0.47 	&	0.04 	&	0.04 	&	0.07 	&	0.07 	&	0.05 	&	7.15 	$\pm$	0.02 	& $\pm$	0.01 	\\
&	$2~^{2}\Pi$	&	-0.08 	&	0.04 	&	0.04 	&	0.05 	&	0.05 	&	0.05 	&	3.47 	$\pm$	0.01 	& $\pm$	0.00 	\\
&	$1~^{4}\Pi$	&	-0.20 	&	-0.03 	&	-0.03 	&	-0.01 	&	0.01 	&	0.02 	&	2.74 	$\pm$	0.00 	& $\pm$	0.00 	\\
CO$^{+}$	&	$2~^{2}\Sigma^{+}$	&	0.83 	&	0.05 	&	0.05 	&	0.07 	&	0.01 	&	-0.01 	&	5.80 	$\pm$	0.00 	& $\pm$	0.00 	\\
&	$1~^{2}\Pi$	&	0.10 	&	0.03 	&	0.04 	&	0.05 	&	-0.01 	&	0.00 	&	3.27 	$\pm$	0.00 	& $\pm$	0.00 	\\
&	$1~^{4}\Sigma^{+}$	&	-0.27 	&	0.06 	&	0.07 	&	0.08 	&	0.02 	&	0.01 	&	7.27 	$\pm$	0.00 	& $\pm$	0.00 	\\
F$_{2}$BO	&	$1~^{2}A_{1}$	&	0.15 	&	0.04 	&	0.04 	&	0.04 	&	0.01 	&	0.01 	&	2.77 	$\pm$	0.01 	& $\pm$	0.04 	\\
&	$2~^{2}A_{1}$	&	1.25 	&	0.21 	&	0.21 	&	0.23 	&	0.04 	&	0.11 	&	7.79 	$\pm$	0.12 	& $\pm$	0.04 	\\
&	$1~^{2}A_{2}$	&	1.23 	&	0.21 	&	0.21 	&	0.22 	&	-0.04 	&	0.07 	&	6.94 	$\pm$	0.07 	& $\pm$	0.03 	\\
&	$2~^{2}B_{2}$	&	1.37 	&	0.26 	&	0.26 	&	0.27 	&	-0.02 	&	0.08 	&	6.53 	$\pm$	0.11 	& $\pm$	0.03 	\\
&	$1~^{2}B_{1}$	&	0.07 	&	-0.01 	&	-0.01 	&	-0.01 	&	-0.01 	&	0.00 	&	0.70 	$\pm$	0.04 	& $\pm$	0.03 	\\
F$_{2}$BS	&	$1~^{2}A_{1}$	&	0.09 	&	0.09 	&	0.09 	&	0.10 	&	-0.01 	&	0.04 	&	2.95 	$\pm$	0.02 	& $\pm$	0.05 	\\
&	$2~^{2}B_{2}$	&	0.09 	&	0.08 	&	0.08 	&	0.10 	&	0.16 	&	0.12 	&	7.02 	$\pm$	0.05 	& $\pm$	0.04 	\\
&	$1~^{2}B_{1}$	&	-0.04 	&	0.05 	&	0.05 	&	0.05 	&	-0.01 	&	-0.02 	&	0.50 	$\pm$	0.00 	& $\pm$	0.04 	\\
&	$1~^{4}B_{2}$	&	-0.03 	&	-0.04 	&	-0.04 	&	-0.03 	&	0.08 	&	0.05 	&	6.39 	$\pm$	0.03 	& $\pm$	0.05 	\\
H$_{2}$BO	&	$1~^{2}A_{1}$	&	0.02 	&	0.06 	&	0.06 	&	0.07 	&	0.01 	&	0.00 	&	3.51 	$\pm$	0.00 	& $\pm$	0.01 	\\
&	$2~^{2}A_{1}$	&	1.16 	&	0.06 	&	0.07 	&	0.09 	&	0.07 	&	0.04 	&	7.25 	$\pm$	0.00 	& $\pm$	0.01 	\\
&	$2~^{2}B_{2}$	&	1.03 	&	0.00 	&	0.00 	&	0.00 	&	0.04 	&	0.02 	&	4.61 	$\pm$	0.00 	& $\pm$	0.01 	\\
&	$3~^{2}B_{2}$	&	1.09 	&	0.05 	&	0.06 	&	0.08 	&	0.00 	&	0.03 	&	7.64 	$\pm$	0.00 	& $\pm$	0.01 	\\
&	$1~^{2}B_{1}$	&	-0.21 	&	0.04 	&	0.04 	&	0.05 	&	-0.02 	&	0.00 	&	2.17 	$\pm$	0.00 	& $\pm$	0.01 	\\
&	$2~^{2}B_{1}$	&	0.98 	&	0.08 	&	0.08 	&	0.11 	&	0.07 	&	0.05 	&	6.13 	$\pm$	0.01 	& $\pm$	0.01 	\\
&	$1~^{4}B_{2}$	&	0.73 	&	0.07 	&	0.07 	&	0.09 	&	0.00 	&	0.05 	&	6.86 	$\pm$	0.00 	& $\pm$	0.01 	\\
HCO	&	$2~^{2}A'$	&	0.24 	&	0.17 	&	0.17 	&	0.17 	&	-0.05 	&	0.03 	&	5.45 	$\pm$	0.01 	& $\pm$	0.01 	\\
&	$3~^{2}A'$	&	0.07 	&	0.17 	&	0.18 	&	0.18 	&	0.07 	&	0.08 	&	6.15 	$\pm$	0.01 	& $\pm$	0.01 	\\
&	$1~^{2}A''$	&	0.17 	&	-0.01 	&	-0.02 	&	0.00 	&	0.04 	&	0.02 	&	2.09 	$\pm$	0.00 	& $\pm$	0.01 	\\
&	$2~^{2}A''$	&	-0.18 	&	0.11 	&	0.11 	&	0.11 	&	-0.02 	&	0.01 	&	7.07 	$\pm$	0.01 	& $\pm$	0.01 	\\
&	$3~^{2}A''$	&	-0.28 	&	0.08 	&	0.10 	&	0.11 	&	-0.01 	&	0.00 	&	7.73 	$\pm$	0.01 	& $\pm$	0.01 	\\
&	$1~^{4}A'$	&	-0.21 	&	0.05 	&	0.05 	&	0.05 	&	-0.02 	&	0.04 	&	6.75 	$\pm$	0.01 	& $\pm$	0.01 	\\
&	$1~^{4}A''$	&	-0.21 	&	0.04 	&	0.05 	&	0.05 	&	0.00 	&	0.05 	&	6.37 	$\pm$	0.01 	& $\pm$	0.01 	\\
HOC	&	$2~^{2}A'$	&	0.36 	&	0.14 	&	0.14 	&	0.13 	&	-0.04 	&	0.02 	&	3.78 	$\pm$	0.00 	& $\pm$	0.01 	\\
&	$3~^{2}A'$	&	0.11 	&	0.13 	&	0.13 	&	0.14 	&	0.09 	&	0.04 	&	5.63 	$\pm$	0.01 	& $\pm$	0.01 	\\
&	$1~^{2}A''$	&	0.01 	&	-0.01 	&	-0.01 	&	0.00 	&	0.02 	&	0.00 	&	0.92 	$\pm$	0.00 	& $\pm$	0.01 	\\
&	$2~^{2}A''$	&	0.20 	&	0.07 	&	0.07 	&	0.10 	&	0.06 	&	0.03 	&	6.13 	$\pm$	0.01 	& $\pm$	0.01 	\\
&	$3~^{2}A''$	&	0.28 	&	0.15 	&	0.15 	&	0.17 	&	0.08 	&	0.06 	&	7.55 	$\pm$	0.01 	& $\pm$	0.01 	\\
&	$1~^{4}A'$	&	0.24 	&	0.18 	&	0.18 	&	0.17 	&	-0.02 	&	0.05 	&	7.73 	$\pm$	0.01 	& $\pm$	0.01 	\\
&	$1~^{4}A''$	&	0.05 	&	-0.08 	&	-0.08 	&	-0.08 	&	0.00 	&	0.02 	&	3.84 	$\pm$	0.01 	& $\pm$	0.01 	\\
H$_{2}$PO	&	$2~^{2}A'$	&	-0.05 	&	0.18 	&	0.24 	&	0.25 	&	0.08 	&	0.04 	&	4.21 	$\pm$	0.01 	& $\pm$	0.03 	\\
&	$3~^{2}A'$	&	0.41 	&	0.08 	&	0.18 	&	0.20 	&	0.06 	&	0.05 	&	4.78 	$\pm$	0.01 	& $\pm$	0.04 	\\
&	$4~^{2}A'$	&	-0.03 	&	0.14 	&	0.21 	&	0.23 	&	0.02 	&	0.03 	&	5.66 	$\pm$	0.02 	& $\pm$	0.02 	\\
&	$1~^{2}A''$	&	-0.01 	&	0.11 	&	0.17 	&	0.16 	&	-0.04 	&	0.01 	&	2.81 	$\pm$	0.01 	& $\pm$	0.03 	\\
&	$2~^{2}A''$	&	0.37 	&	0.63\tnote{d} &	0.41 	&	0.44 	&	0.30 	&	0.21 	&	6.20 	$\pm$	0.02 	& $\pm$	0.04 	\\
&	$3~^{2}A''$	&	0.65 	&	-0.11\tnote{d} &	0.23 	&	0.23 	&	-0.04 	&	0.05 	&	6.81 	$\pm$	0.03 	& $\pm$	0.03 	\\
&	$1~^{4}A'$	&	0.06 	&	0.18 	&	0.24 	&	0.25 	&	0.02 	&	0.10 	&	7.28 	$\pm$	0.02 	& $\pm$	0.03 	\\
&	$2~^{4}A'$	&	0.57 	&	0.27 	&	0.32 	&	0.32 	&	0.10 	&	0.18 	&	7.64 	$\pm$	0.02 	& $\pm$	0.03 	\\
&	$1~^{4}A''$	&	0.53 	&	0.16 	&	0.22 	&	0.23 	&	0.02 	&	0.12 	&	6.31 	$\pm$	0.02 	& $\pm$	0.03 	\\
H$_{2}$PS	&	$2~^{2}A'$	&	0.13 	&	0.09 	&	0.11 	&	0.11 	&	0.04 	&	0.02 	&	2.73 	$\pm$	0.01 	& $\pm$	0.01 	\\
&	$3~^{2}A'$	&	0.12 	&	0.19 	&	0.20 	&	0.21 	&	0.02 	&	0.02 	&	4.57 	$\pm$	0.02 	& $\pm$	0.01 	\\
&	$1~^{2}A''$	&	-0.09 	&	0.08 	&	0.09 	&	0.09 	&	-0.02 	&	0.02 	&	1.14 	$\pm$	0.00 	& $\pm$	0.01 	\\
&	$2~^{2}A''$	&	0.59 	&	0.12 	&	0.13 	&	0.14 	&	0.04 	&	0.09 	&	5.42 	$\pm$	0.03 	& $\pm$	0.01 	\\
&	$1~^{4}A'$	&	0.66 	&	0.17 	&	0.18 	&	0.17 	&	0.03 	&	0.10 	&	5.74 	$\pm$	0.01 	& $\pm$	0.01 	\\
&	$2~^{4}A'$	&	0.68\tnote{d} &	0.23 	&	0.19 	&	0.22 	&	0.07 	&	0.07 	&	6.02 	$\pm$	0.01 	& $\pm$	0.01 	\\
&	$3~^{4}A'$	&	-0.76\tnote{d} &	0.34 	&	0.40 	&	0.40 	&	0.15 	&	0.17 	&	6.76 	$\pm$	0.02 	& $\pm$	0.01 	\\
&	$1~^{4}A''$	&	1.09\tnote{d} 	&	0.21 	&	0.19 	&	0.20 	&	0.10 	&	0.10 	&	5.08 	$\pm$	0.01 	& $\pm$	0.02 	\\
&	$2~^{4}A''$	&	-0.34\tnote{d} 	&	0.46 	&	0.49 	&	0.51 	&	0.23 	&	0.22 	&	5.73 	$\pm$	0.02 	& $\pm$	0.01 	\\
&	$3~^{4}A''$	&	-0.05 	&	0.14 	&	0.16 	&	0.16 	&	-0.02 	&	0.06 	&	6.72 	$\pm$	0.01 	& $\pm$	0.01 	\\
NCO	&	$1~^{2}\Sigma^{+}$	&	0.22 	&	-0.04 	&	-0.03 	&	-0.03 	&	0.05 	&	0.01 	&	2.87 	$\pm$	0.01 	& $\pm$	0.03 	\\
&	$2~^{2}\Sigma^{+}$	&	1.22 	&	0.07 	&	0.08 	&	0.08 	&	0.03 	&	0.03 	&	7.03 	$\pm$	0.02 	& $\pm$	0.02 	\\
&	$2~^{2}\Pi$	&	0.88 	&	0.03 	&	0.05 	&	0.05 	&	0.02 	&	0.02 	&	4.72 	$\pm$	0.01 	& $\pm$	0.02 	\\
&	$3~^{2}\Pi$	&	0.04 	&	0.10 	&	0.11 	&	0.12 	&	0.07 	&	0.05 	&	7.43 	$\pm$	0.03 	& $\pm$	0.02 	\\
&	$1~^{4}\Pi$	&	-0.12 	&	0.06 	&	0.07 	&	0.08 	&	0.04 	&	0.06 	&	6.79 	$\pm$	0.01 	& $\pm$	0.02 	\\
NH$_{2}$	&	$1~^{2}A_{1}$	&	0.08 	&	0.02 	&	0.02 	&	0.02 	&	0.00 	&	0.00 	&	2.12 	$\pm$	0.00 	& $\pm$	0.00 	\\
&	$2~^{2}A_{1}$	&	0.04 	&	0.06 	&	0.06 	&	0.05 	&	0.00 	&	0.00 	&	7.69 	$\pm$	0.00 	& $\pm$	0.00 	\\
&	$1~^{2}B_{2}$	&	0.01 	&	0.06 	&	0.06 	&	0.07 	&	0.01 	&	0.01 	&	6.48 	$\pm$	0.00 	& $\pm$	0.00 	\\
&	$2~^{2}B_{1}$	&	0.02 	&	0.03 	&	0.03 	&	0.02 	&	-0.02 	&	0.00 	&	7.75 	$\pm$	0.00 	& $\pm$	0.00 	\\
&	$1~^{4}B_{1}$	&	0.04 	&	0.04 	&	0.04 	&	0.03 	&	0.00 	&	0.01 	&	7.29 	$\pm$	0.00 	& $\pm$	0.00 	\\
Nitromethyl  	&	$1~^{2}A_{1}$	&	0.84 	&	-0.03 	&	0.03 	&	-0.01 	&	0.21 	&	0.15 	&	2.50 	$\pm$	0.01 	& $\pm$	0.03 	\\
(CH$_{2}$NO$_{2}$)	&	$2~^{2}A_{1}$	&	0.35 	&	-0.05 	&	0.01 	&	-0.01 	&	0.06 	&	0.07 	&	5.20 	$\pm$	0.01 	& $\pm$	0.04 	\\
&	$1~^{2}A_{2}$	&	1.13 	&	-0.05 	&	-0.02 	&	-0.04 	&	0.10 	&	0.08 	&	2.30 	$\pm$	0.00 	& $\pm$	0.03 	\\
&	$2~^{2}A_{2}$	&	0.64 	&	0.03 	&	0.13 	&	0.15 	&	0.11 	&	0.15 	&	6.70 	$\pm$	0.07 	& $\pm$	0.03 	\\
&	$1~^{2}B_{2}$	&	1.02 	&	-0.03 	&	0.03 	&	0.01 	&	0.13 	&	0.11 	&	2.00 	$\pm$	0.02 	& $\pm$	0.04 	\\
&	$2~^{2}B_{2}$	&	0.33 	&	-0.02 	&	0.04 	&	0.02 	&	-0.01 	&	0.03 	&	4.66 	$\pm$	0.02 	& $\pm$	0.04 	\\
&	$2~^{2}B_{1}$	&	0.21 	&	-0.05 	&	0.08 	&	0.13 	&	0.42 	&	0.30 	&	5.31 	$\pm$	0.00 	& $\pm$	0.03 	\\
&	$1~^{4}A_{1}$	&	0.04 	&	0.00 	&	0.06 	&	0.04 	&	0.06 	&	0.09 	&	4.94 	$\pm$	0.03 	& $\pm$	0.02 	\\
&	$1~^{4}A_{2}$	&	-0.52 	&	-0.05 	&	0.01 	&	-0.02 	&	-0.05 	&	0.00 	&	4.30 	$\pm$	0.02 	& $\pm$	0.03 	\\
&	$1~^{4}B_{2}$	&	0.10 	&	0.01 	&	0.07 	&	0.05 	&	0.00 	&	0.05 	&	4.46 	$\pm$	0.03 	& $\pm$	0.04 	\\
NO	&	$1~^{2}\Sigma^{+}$	&	0.46 	&	0.19 	&	0.19 	&	0.19 	&	-0.09 	&	-0.01 	&	6.11 	$\pm$	0.01 	& $\pm$	0.01 	\\
&	$2~^{2}\Sigma^{+}$	&	0.28 	&	0.17 	&	0.17 	&	0.16 	&	-0.11 	&	-0.02 	&	7.28 	$\pm$	0.01 	& $\pm$	0.00 	\\
&	$2~^{2}\Pi$	&	-0.45 	&	-0.04 	&	-0.04 	&	-0.04 	&	-0.15 	&	-0.13 	&	7.79 	$\pm$	0.00 	& $\pm$	0.01 	\\
&	$1~^{4}\Sigma^{-}$	&	-0.21 	&	-0.04 	&	-0.04 	&	-0.02 	&	0.05 	&	0.02 	&	6.36 	$\pm$	0.00 	& $\pm$	0.01 	\\
&	$1~^{4}\Pi$	&	-0.45 	&	0.06 	&	0.06 	&	0.05 	&	-0.03 	&	0.02 	&	6.78 	$\pm$	0.00 	& $\pm$	0.01 	\\
OH	&	$1~^{2}\Sigma^{+}$	&	0.04 	&	0.01 	&	0.01 	&	0.01 	&	0.01 	&	0.00 	&	4.11 	$\pm$	0.00 	& $\pm$	0.00 	\\
&	$1~^{2}\Sigma^{-}$	&	0.09 	&	0.01 	&	0.01 	&	0.01 	&	-0.01 	&	-0.01 	&	8.03 	$\pm$	0.00 	& $\pm$	0.00 	\\
&	$1~^{4}\Sigma^{-}$	&	0.09 	&	0.02 	&	0.02 	&	0.02 	&	0.00 	&	-0.01 	&	7.50 	$\pm$	0.00 	& $\pm$	0.00 	\\
PH$_{2}$	&	$1~^{2}A_{1}$	&	0.05 	&	0.07 	&	0.07 	&	0.07 	&	0.04 	&	-0.01 	&	2.76 	$\pm$	0.00 	& $\pm$	0.00 	\\
&	$2~^{2}A_{1}$	&	-0.15 	&	0.05 	&	0.05 	&	0.04 	&	0.00 	&	0.02 	&	5.90 	$\pm$	0.00 	& $\pm$	0.00 	\\
&	$1~^{2}A_{2}$	&	0.24 	&	0.21 	&	0.21 	&	0.22 	&	0.10 	&	0.02 	&	7.10 	$\pm$	0.00 	& $\pm$	0.00 	\\
&	$1~^{2}B_{2}$	&	0.08 	&	0.10 	&	0.10 	&	0.10 	&	0.03 	&	-0.02 	&	5.17 	$\pm$	0.00 	& $\pm$	0.00 	\\
&	$2~^{2}B_{2}$	&	-0.07 	&	0.08 	&	0.09 	&	0.09 	&	0.02 	&	-0.02 	&	5.75 	$\pm$	0.00 	& $\pm$	0.00 	\\
&	$2~^{2}B_{1}$	&	-0.14 	&	0.06 	&	0.06 	&	0.06 	&	0.01 	&	0.01 	&	7.47 	$\pm$	0.00 	& $\pm$	0.00 	\\
&	$1~^{4}A_{2}$	&	0.09 	&	0.04 	&	0.04 	&	0.04 	&	0.00 	&	0.03 	&	6.16 	$\pm$	0.00 	& $\pm$	0.00 	\\
&	$1~^{4}B_{1}$	&	-0.03 	&	0.13 	&	0.13 	&	0.13 	&	0.06 	&	0.05 	&	6.95 	$\pm$	0.00 	& $\pm$	0.00 	\\
Vinyl  	&	$2~^{2}A'$	&	0.75\tnote{d}  	&	0.10 	&	0.09 	&	0.11 	&	0.02 	&	0.02 	&	5.60 	$\pm$	0.00 	& $\pm$	0.02 	\\
(C$_{2}$H$_{3}$)	&	$3~^{2}A'$	&	-0.80\tnote{d} 	&	0.13 	&	0.17 	&	0.15 	&	-0.06 	&	0.01 	&	6.19 	$\pm$	0.01 	& $\pm$	0.01 	\\
&	$1~^{2}A''$	&	-0.10 	&	0.04 	&	0.05 	&	0.05 	&	0.01 	&	-0.03 	&	3.28 	$\pm$	0.00 	& $\pm$	0.01 	\\
&	$2~^{2}A''$	&	-0.03 	&	0.05 	&	0.06 	&	0.09 	&	0.08 	&	0.02 	&	4.70 	$\pm$	0.00 	& $\pm$	0.01 	\\
&	$3~^{2}A''$	&	-0.42 	&	0.14 	&	0.16 	&	0.15 	&	-0.07 	&	0.02 	&	7.54 	$\pm$	0.01 	& $\pm$	0.01 	\\
&	$1~^{4}A'$	&	-0.15 	&	0.05 	&	0.06 	&	0.05 	&	-0.03 	&	0.02 	&	4.54 	$\pm$	0.01 	& $\pm$	0.01 	\\
&	$1~^{4}A''$	&	-0.38 	&	0.16 	&	0.17 	&	0.15 	&	-0.08 	&	0.03 	&	7.40 	$\pm$	0.01 	& $\pm$	0.01 	\\

			\end{longtable}
			\begin{tablenotes}
%				\item[a] $\pm x$ means extrapolation distance $|x|$.                                     
%				\textsuperscript{b}Sum of the 95\% confidence intervals of ground and excited states. 
%				\textsuperscript{c}The AVTZ basis set of Be varies across different program packages. 
%				\textsuperscript{d} Incorrect ordering.
				\item[a] $\pm x$ means extrapolation distance $|x|$.              
				\item[b] Sum of the 95\% confidence intervals of ground and excited states.
				\item[c] The AVTZ basis set of Be varies across different program packages.
				\item[d] Incorrect ordering.
			\end{tablenotes}
		\end{ThreePartTable}
	\end{center}
}

{																							
	\setlength{\tabcolsep}{6pt}
	\scriptsize
	\begin{center}																					
		\begin{threeparttable}[]																			
\caption{Statistical analysis of the errors (in eV) of X-TDA, SS-NEVPT2, MS-NEVPT, SDSPT2, SDSCI, and ic-MRCISD relative to iCIPT2 for
	the vertical excitation energies of 110 doublet and 39 quartet states$^\ast$}																
			\label{table-wrtici}													
			\begin{tabular}{clcccccc}																							
\toprule[0.5pt]										
Type&	&X-TDA & SS-NEVPT2 & MS-NEVPT2 & SDSPT2 & SDSCI &	ic-MRCISD\\
\toprule[0.5pt]
Doublets	&	MD	&	0.17 	&	0.07 	&	0.08 	&	0.08 	&	0.02 	&	0.02 	\\
(110 states)	&	MAD	&	0.40 	&	0.10 	&	0.09 	&	0.10 	&	0.06 	&	0.04 	\\
&	SD	&	0.58 	&	0.15 	&	0.12 	&	0.12 	&	0.08 	&	0.06 	\\
&	MAX	&	2.42 	&	0.63 	&	0.41 	&	0.44 	&	0.42 	&	0.30 	\\
\toprule[0.5pt]	
Quartets	&	MD	&	-0.03 	&	0.09 	&	0.11 	&	0.11 	&	0.01 	&	0.04 	\\
(39 states)	&	MAD	&	0.31 	&	0.11 	&	0.12 	&	0.12 	&	0.05 	&	0.06 	\\
&	SD	&	0.41 	&	0.15 	&	0.16 	&	0.16 	&	0.08 	&	0.08 	\\
&	MAX	&	1.09 	&	0.46 	&	0.49 	&	0.51 	&	0.23 	&	0.22 	\\
\toprule[0.5pt]	
Overall&MD	&	0.12 	&	0.07 	&	0.09 	&	0.09 	&	0.02 	&	0.03 	\\
(149 states) &MAD	&	0.38 	&	0.10 	&	0.10 	&	0.10 	&	0.05 	&	0.04 	\\
&SD	&	0.54 	&	0.15 	&	0.13 	&	0.13 	&	0.08 	&	0.07 	\\
&MAX	&	2.42 	&	0.63 	&	0.49 	&	0.51 	&	0.42 	&	0.30 	\\

\bottomrule[0.5pt]											
				
			\end{tabular}
\begin{tablenotes}
	\item[*]MD: mean deviation; MAD: mean absolute deviation; SD: standard deviation; MAX: maximum deviation.
\end{tablenotes}																			
		\end{threeparttable}
	\end{center}																						
}

{																							
	\setlength{\tabcolsep}{6pt}
	\scriptsize
	\begin{center}																					
		\begin{threeparttable}[]																			
			\caption{Internal consistency (percentage number of states within a deviation range in absolute value) across SS-NEVPT2, MS-NEVPT2, SDSPT2, SDSCI, ic-MRCISD, and iCIPT2}	
			\label{table-devrange}													
			\begin{tabular}{cccccccc}
				\toprule[0.5pt]
				\textbf{\makecell[c]{Deviation \\Range (eV)}}& \textbf{\makecell[c]{SS-NEVPT2\\vs\\MS-NEVPT2}}& \textbf{\makecell[c]{SS-NEVPT2\\vs\\SDSPT2}}&
				\textbf{\makecell[c]{MS-NEVPT2\\vs\\SDSPT2}}&\textbf{\makecell[c]{MS-NEVPT2\\vs\\SDSCI}}&\textbf{\makecell[c]{SDSPT2\\vs\\SDSCI}}&\textbf{\makecell[c]{SDSCI\\vs\\ic-MRCISD}}&\textbf{\makecell[c]{ic-MRCISD\\vs\\iCIPT2}}\\
				\toprule[0.5pt]
%0.02 	&	81 	\% &	75 \% &	90 \% &	22 \% &	22 \% &	49 \% &	52\% 	\\
0.05 	&	84 	\% &	89 \% &	99 \% &	50 \% &	44 \% &	75 \% &	74\% 	\\
0.10 	&	95 	\% &	95 \% &	99 \% &	72 \% &	68 \% &	95 \% &	89\% 	\\
0.15 	&	97 	\% &	96 \% &100 \% &82 \% &	81 \% &	100 \% &96\% 	\\

				\bottomrule[0.5pt]											
				
			\end{tabular}
%			\begin{tablenotes}
%				\item[*]$\Delta$MRPT2: difference between the maximum and minimum values among SDSPT2, MS-NEVPT2, and SS-NEVPT2 for each state.
%			\end{tablenotes}																			
		\end{threeparttable}
	\end{center}																						
}	
%%%%%%%%%%%%%%%%%%%%%%%%%%%%%%%%%%%%%%%%%%%%%%%%%%%%%%%%%%%%%%%%%%%%%%%%%%%%%%%%%%%%%%%%%%%%%%%%%%%%%%

{																							
	\setlength{\tabcolsep}{6pt}
	\scriptsize
	\begin{center}																					
		\begin{threeparttable}[]																			
			\caption{CASSCF, MS-NEVPT2, and SDSPT2 wavefunctions of excited states of H$_{2}$PO and CH$_{3}$ }						
			\label{table-H2PO}													
			\begin{tabular}{lcc}																	
				\toprule[0.5pt]	
				H$_{2}$PO& \multicolumn{2}{l}{Ground state: (10a')$^{1}$(3a'')$^{2}$}\\	
				\toprule[0.5pt]									
				&	(3a'')$^{2}$ $\to$ (11a')$^{0}$	&	(10a')$^{1}$ $\to$ (4a'')$^{1}$ \\
				$\Psi^{\mathrm{CASSCF}}_{2~^{2}A''}	$&	\textbf{72\%}	&	15\%\\
				$\Psi^{\mathrm{CASSCF}}_{3~^{2}A''}	$&	18\%	&	\textbf{62\%}\\
				$\Psi^{\mathrm{MS-NEVPT2}}_{2~^{2}A''}	$&	3\%	&	\textbf{75\%}\\
				$\Psi^{\mathrm{MS-NEVPT2}}_{3~^{2}A''}	$&	\textbf{87\%}	&	3\%\\
				$\Psi^{\mathrm{SDSPT2}}_{2~^{2}A''}	$&	7\%	&	\textbf{73\%}\\
				$\Psi^{\mathrm{SDSPT2}}_{3~^{2}A''}	$&\textbf{	84\%}	&	5\%\\
				\toprule[0.5pt]	
				CH$_{3}$&\multicolumn{2}{l}{Ground state: (3a$_1$)$^2$(1b$_2$)$^2$(1b$_1$)$^1$}\\	
				\toprule[0.5pt]	
				&(1b$_1$)$^1$ $\to$ (5a$_1$)$^0$&	(3a$_1$)$^2$ $\to$ (1b$_1$)$^1$\\
				$\Psi^{\mathrm{CASSCF}}_{2~^{2}A_{1}(1~^{2}E')}	$&	\textbf{96\%}	&	0\%\\
				$\Psi^{\mathrm{CASSCF}}_{3~^{2}A_{1}(2~^{2}E')}	$&	0\%	&	\textbf{95\%}\\
				$\Psi^{\mathrm{MS-NEVPT2}}_{2~^{2}A_{1}(1~^{2}E')}	$&	20\%	&	\textbf{76\%}\\
				$\Psi^{\mathrm{MS-NEVPT2}}_{3~^{2}A_{1}(2~^{2}E')}	$&	\textbf{75\%}	&	20\%\\
				$\Psi^{\mathrm{SDSPT2}}_{2~^{2}A_{1}(1~^{2}E')}	$&	\textbf{42\%}	&	\textbf{54\%}\\
				$\Psi^{\mathrm{SDSPT2}}_{3~^{2}A_{1}(2~^{2}E')}	$&\textbf{54\%}	&	\textbf{41\%}\\
				&(1b$_1$)$^1$ $\to$ (2b$_2$)$^0$&	(1b$_2$)$^2$ $\to$ (1b$_1$)$^1$\\
				$\Psi^{\mathrm{CASSCF}}_{1~^{2}B_{2}(1~^{2}E')}	$&	\textbf{96\%}	&	0\%\\
				$\Psi^{\mathrm{CASSCF}}_{2~^{2}B_{2}(2~^{2}E')}	$&	0\%	&	\textbf{95\%}\\
				$\Psi^{\mathrm{MS-NEVPT2}}_{1~^{2}B_{2}(1~^{2}E')}$&	20\%	&	\textbf{76\%}\\
				$\Psi^{\mathrm{MS-NEVPT2}}_{2~^{2}B_{2}(2~^{2}E')}$&	\textbf{75\%}	&	20\%\\
				$\Psi^{\mathrm{SDSPT2}}_{1~^{2}B_{2}(1~^{2}E')}	$&	\textbf{48\%}	&	\textbf{48\%}\\
				$\Psi^{\mathrm{SDSPT2}}_{2~^{2}B_{2}(2~^{2}E')}	$&\textbf{48\%}	&	\textbf{47\%}\\
				
				\bottomrule[0.5pt]											
				
			\end{tabular}
			%			\begin{tablenotes}
				%				\item[*]H$_{2}$PO: $2~^{2}A''$ and $3~^{2}A''$; CH$_{3}$: $1~^{2}E'$ ($1~^{2}B_{1}+2~^{2}A_{1}$) and $2~^{2}E'$ ($2~^{2}B_{1}+3~^{2}A_{1}$)
				%			\end{tablenotes}																			
		\end{threeparttable}
	\end{center}																						
}

{																							
	\setlength{\tabcolsep}{6pt}
	\scriptsize
	\begin{center}																					
		\begin{threeparttable}[]																			
			\caption{Vertical excitation energies (in eV) of C$_{3}$H$_{5}$ }						
			\label{table-C3H5-ene}													
			\begin{tabular}{lcccccccc}	
\toprule[0.5pt]	
	&	\textbf{Type}\tnote{a}	&	\textbf{SS-NEVPT2}	&	\textbf{MS-NEVPT2}	&	\textbf{SDSPT2}	&	\textbf{SDSCI}	&	\textbf{ic-MRCISD}	&	\textbf{iCIPT2}\tnote{b} &\textbf{95\% interval}\tnote{c}	\\				
	&			&	\textbf{AVTZ} 		&	\textbf{AVTZ}		&	\textbf{AVTZ}	&	\textbf{AVTZ}	&	\textbf{AVTZ}		&	\textbf{AVTZ}	&		\\
				\toprule[0.5pt]	
$1~^{2}B_{1}$	&	Val.	&	3.57 	&	3.56 	&	3.50 	&	3.28 	&	3.36 	&	3.40 	$\pm$	0.02	& $\pm$	0.03	\\
$1~^{2}A_{1}$	&	Ryd.	&	5.13 	&	5.13 	&	5.10 	&	4.79 	&	4.87 	&	4.94 	$\pm$	0.04	& $\pm$	0.03	\\
$2~^{2}A_{1}$	&	Ryd.	&	5.70 	&	5.70 	&	5.68 	&	5.35 	&	5.44 	&	5.46 	$\pm$	0.06	& $\pm$	0.03	\\
$1~^{2}B_{2}$	&	Ryd.	&	5.79 	&	5.79 	&	5.77 	&	5.43 	&	5.53 	&	5.61 	$\pm$	0.01	& $\pm$	0.03	\\
$2~^{2}B_{1}$	&	Ryd.	&	6.12 	&	5.78 	&	5.93 	&	5.74 	&	5.79 	&	5.71 	$\pm$	0.05	& $\pm$	0.02	\\
$1~^{4}A_{2}$	&	Val.	&	6.20 	&	6.20 	&	6.17 	&	5.73 	&	5.86 	&	5.98 	$\pm$	0.02	& $\pm$	0.02	\\
$3~^{2}B_{1}$	&	Ryd.	&	7.00 	&	6.65 	&	6.68 	&	6.60 	&	6.60 	&	6.53 	$\pm$	0.01	& $\pm$	0.03	\\
$1~^{4}B_{2}$	&	Ryd.	&	7.92 	&	7.92 	&	7.89 	&	7.41 	&	7.54 	&	7.68 	$\pm$	0.02	& $\pm$	0.02	\\
$4~^{2}B_{1}$	&	Ryd.	&	7.17 	&	7.87 	&	7.73 	&	7.72 	&	7.69 	&	7.80 	$\pm$	0.02	& $\pm$	0.04	\\
\toprule[0.5pt]	
&	\textbf{Type}\tnote{a}	&	\textbf{CI2}\tnote{d}	&	\textbf{ic-MRCISD}\tnote{e}	&	\textbf{MS-CASPT2}\tnote{f}	&	\textbf{ROCC3}\tnote{h}	&	\textbf{mcf effect}\tnote{i}	&	\textbf{iCIPT2}\tnote{b}			&\textbf{95\% interval}\tnote{c}		\\
&		&	\textbf{DZP}			&	\textbf{DZP}			&	\textbf{ANO-L}		&	\textbf{AVTZ} 	&	\textbf{of UCCSD}	&	\textbf{AVTZ}			&		\\
&		&	\textbf{+ 2s1p on C}	& \textbf{+ 1s1p on C}	&\textbf{+ 1s1p1d on C}	& \textbf{+ mcf: 2s2p2d}	&		&				&		\\
\toprule[0.5pt]	
$1~^{2}B_{1}$	&	Val.	&	3.36 	&	3.32	&	3.32 	&	3.44 	&	0.03 	&	3.40 	$\pm$	0.02	& $\pm$	0.03	\\
$1~^{2}A_{1}$	&	Ryd.	&	4.70 	&	4.68	&	5.11 	&	4.94 	&	-0.01 	&	4.94 	$\pm$	0.04	& $\pm$	0.03	\\
$2~^{2}A_{1}$	&	Ryd.	&	5.25 	&	5.25	&	5.65 	&	5.49 	&	-0.02 	&	5.46 	$\pm$	0.06	& $\pm$	0.03	\\
$1~^{2}B_{2}$	&	Ryd.	&	5.31 	&	5.29	&	5.76 	&	5.59 	&	0.09 	&	5.61 	$\pm$	0.01	& $\pm$	0.03	\\
$2~^{2}B_{1}$	&	Ryd.	&	5.50 	&	5.46	&	5.73 	&	5.65 	&	-0.09 	&	5.71 	$\pm$	0.05	& $\pm$	0.02	\\
$1~^{4}A_{2}$	&	Val.	&	- &	-	&	5.89\tnote{g}	&	6.15 	&	0.04 	&	5.98 	$\pm$	0.02	& $\pm$	0.02	\\
$3~^{2}B_{1}$	&	Ryd.	&	-	&	-	&	6.36 	&	6.20 	&	-0.49 	&	6.53 	$\pm$	0.01	& $\pm$	0.03	\\
$1~^{4}B_{2}$	&	Ryd.	&	-	&	-	&	-	&	-	&	-	&	7.68 	$\pm$	0.02	& $\pm$	0.02	\\
$4~^{2}B_{1}$	&	Ryd.	&	-	&	-	&	6.90 	&	6.79 	&	-1.44 	&	7.80 	$\pm$	0.02	& $\pm$	0.04	\\
\bottomrule[0.5pt]											

\end{tabular}
\begin{tablenotes}
\item[a] Val.: valence excitation; Ryd.: Rydberg excitation.
\item[b] $\pm x$ means extrapolation distance $|x|$.
\item[c] Sum of the 95\% confidence intervals of ground and excited states.
\item[d] Ref. \citenum{C3H5_MRCI_03JACS}; second-order MRCI with (3e,7o).
\item[e] Ref. \citenum{C3H5_MRCI_10JPCA}; ic-MRCISD with (3e,7o).
\item[f] Ref. \citenum{C3H5_CASPT2}; ANO-L: C(14s9p4d)/H(8s4p) $\to$ C[4s3p2d]/H[3s2p]; (3e,7o) for the doublets, whereas (3e,3o) for $1~^{4}A_{2}$.
\item[g] Ref. \citenum{C3H5_CASPT2}; CASPT2 with (3e,3o).
\item[h] Ref. \citenum{C3H5_CC_2010}; mcf:  molecule-centered basis functions. 
\item[i] Ref. \citenum{C3H5_CC_2010}; E(UCCSD/AVTZ + mcf) $-$ E(UCCSD/AVTZ).
\end{tablenotes}
\end{threeparttable}
\end{center}																						
}

{																							
	\setlength{\tabcolsep}{6pt}
	\scriptsize
	\begin{center}																					
		\begin{threeparttable}[]																			
			\caption{CASSCF, MS-NEVPT2, SDSPT2, SDSCI, and ic-MRCISD wavefunctions of excited states of C$_{3}$H$_{5}$}						
			\label{table-C3H5}													
			\begin{tabular}{lccccc}	
				\toprule[0.5pt]	
				C$_{3}$H$_{5}$& \multicolumn{5}{l}{Ground state: (6a$_1$)$^2$(1a$_2$)$^1$(4b$_2$)$^2$(1b$_1$)$^2$}\\					
				\toprule[0.5pt]	
				&(1a$_2$)$^1$ $\to$ (2b$_1$)$^0$	&	(1b$_1$)$^2$ $\to$ (1a$_2$)$^1$	&	(1a$_2$)$^1$ $\to$ (3b$_1$)$^0$	&	(1a$_2$)$^1$ $\to$ (4b$_1$)$^0$&Rydberg\\
				&Valence & Valence & Rydberg &Rydberg&component \\
				\toprule[0.5pt]											
				$\Psi^{\mathrm{CASSCF}}_{1~^{2}B_{1}}$	&	41\%	&	49\%	&	1\%	&	0\%	&	1\%	\\
				$\Psi^{\mathrm{CASSCF}}_{2~^{2}B_{1}}$	&	10\%	&	5\%	&	51\%	&	28\%	&	79\%	\\
				$\Psi^{\mathrm{CASSCF}}_{3~^{2}B_{1}}$	&	8\%	&	11\%	&	41\%	&	33\%	&	74\%	\\
				$\Psi^{\mathrm{CASSCF}}_{4~^{2}B_{1}}$	&	29\%	&	25\%	&	0\%	&	32\%	&	32\%	\\
				\toprule[0.5pt]											
				$\Psi^{\mathrm{MS-NEVPT2}}_{1~^{2}B_{1}}$	&	43\%	&	48\%	&	0\%	&	0\%	&	0\%	\\
				$\Psi^{\mathrm{MS-NEVPT2}}_{2~^{2}B_{1}}$	&	37\%	&	30\%	&	11\%	&	14\%	&	25\%	\\
				$\Psi^{\mathrm{MS-NEVPT2}}_{3~^{2}B_{1}}$	&	6\%	&	10\%	&	71\%	&	3\%	&	74\%	\\
				$\Psi^{\mathrm{MS-NEVPT2}}_{4~^{2}B_{1}}$	&	2\%	&	3\%	&	10\%	&	76\%	&	86\%	\\
				\toprule[0.5pt]											
				$\Psi^{\mathrm{SDSPT2}}_{1~^{2}B_{1}}$	&	41\%	&	49\%	&	0\%	&	0\%	&	0\%	\\
				$\Psi^{\mathrm{SDSPT2}}_{2~^{2}B_{1}}$	&	21\%	&	13\%	&	34\%	&	26\%	&	60\%	\\
				$\Psi^{\mathrm{SDSPT2}}_{3~^{2}B_{1}}$	&	14\%	&	17\%	&	55\%	&	4\%	&	59\%	\\
				$\Psi^{\mathrm{SDSPT2}}_{4~^{2}B_{1}}$	&	11\%	&	12\%	&	3\%	&	63\%	&	66\%	\\
				\toprule[0.5pt]											
				$\Psi^{\mathrm{SDSCI}}_{1~^{2}B_{1}}$	&	40\%	&	51\%	&	0\%	&	0\%	&	0\%	\\
				$\Psi^{\mathrm{SDSCI}}_{2~^{2}B_{1}}$	&	18\%	&	9\%	&	39\%	&	28\%	&	67\%	\\
				$\Psi^{\mathrm{SDSCI}}_{3~^{2}B_{1}}$	&	14\%	&	15\%	&	52\%	&	11\%	&	63\%	\\
				$\Psi^{\mathrm{SDSCI}}_{4~^{2}B_{1}}$	&	17\%	&	16\%	&	2\%	&	54\%	&	56\%	\\
				\toprule[0.5pt]											
				$\Psi^{\mathrm{ic-MRCISD}}_{1~^{2}B_{1}}$	&	36\%	&	45\%	&	0\%	&	0\%	&	0\%	\\
				$\Psi^{\mathrm{ic-MRCISD}}_{2~^{2}B_{1}}$	&	18\%	&	11\%	&	32\%	&	23\%	&	55\%	\\
				$\Psi^{\mathrm{ic-MRCISD}}_{3~^{2}B_{1}}$	&	13\%	&	15\%	&	48\%	&	5\%	&	53\%	\\
				$\Psi^{\mathrm{ic-MRCISD}}_{4~^{2}B_{1}}$	&	11\%	&	11\%	&	3\%	&	55\%	&	58\%	\\
					\bottomrule[0.5pt]											
				
			\end{tabular}
			
		\end{threeparttable}
	\end{center}																						
}

\section{Conclusions}\label{Conclusion}
A comprehensive dataset, QUEST\#4X, has been introduced for the purpose of calibrating multireference (MR) methods.
It encompasses 110 doublet and 39 quartet excited states for the 24 radicals in the dataset QUEST\#4 (which has only 51 doublet states).
The iCIPT2/AVTZ vertical excitation energies reported here serve as a benchmark for any other MR methods. It is of interest to note
that the MR methods considered here, i.e., NEVPT2, SDSPT2, SDSCI, and ic-MRCISD, share the same function space and differ only in the determination of
the expansion coefficients. Because of this, SDSCI can be taken as the initial step of ic-MRCISD and can even yield SDSPT2 and MS-NEVPT2 results automatically.
Moreover, SDSPT2 also yields MS-NEVPT2 results automatically. More interestingly,
the degree of agreement between SDSPT2 and MS-NEVPT2, the magnitude of deviations of SDSPT2/MS-NEVPT2 from SDSCI,
as well as the iterative increments going from SDSCI to ic-MRCISD are clear indicators
for the internal consistency between the methods and hence their accuracy. A good internal consistency can only be achieved by choosing appropriate
active spaces for all target states, as done here.
Notwithstanding this, more challenging datasets for the excited states of open-shell transition metals and $f$-elements
remain to be established. Work along this direction is being carried out at our laboratory.

\section*{Acknowledgments}
This work was supported by the National Natural Science Foundation of China (Grant No. 22373057) and
Mount Tai Scholar Climbing Project of Shandong Province.

\section*{Supporting Information}
Information for the radicals under concern, proper choice of zeroth-order Hamiltonian for Rydberg states,
valence atomic orbitals, and CASSCF energies. 

\section*{Conflicts of interest}
There are no conflicts to declare.
%\printbibliography
\bibliography{radical}
\end{document}